\documentclass[
    nolinenumbers,
    trackchanges]{aastex701}

\usepackage{txfonts} 
\usepackage{color}
\usepackage{graphicx}
\usepackage{subcaption}
\usepackage{float}
\usepackage{threeparttable}
\usepackage{textcomp}
\usepackage{multirow}
\usepackage{makecell}
\usepackage{hyperref}

\shorttitle{Spectroscopic and Photometric Study of V2790 Ori}
\shortauthors{Wang et al.}

\usepackage[T1]{fontenc}
\begin{document}

\title{Spectroscopic and Decade-long Photometric Observations of the Contact Binary V2790~Ori: Evidence for a Brown Dwarf Companion and a Solar-like Magnetic Activity Cycle}

\correspondingauthor{Kai Li}
\email{kaili@sdu.edu.cn}
\affiliation{Shandong Key Laboratory of Space Environment and Exploration Technology, Institute of Space Sciences, School of Space Science and Technology, Shandong University, Shandong, China.}

\author{Si-Rui Wang}
\email{202317772@mail.sdu.edu.cn}
\affiliation{Shandong Key Laboratory of Space Environment and Exploration Technology, Institute of Space Sciences, School of Space Science and Technology, Shandong University, Shandong, China.}

\author{Kai Li}
\email{kaili@sdu.edu.cn}
\affiliation{Shandong Key Laboratory of Space Environment and Exploration Technology, Institute of Space Sciences, School of Space Science and Technology, Shandong University, Shandong, China.}

\author{Qi-Qi Xia}
\email{xiaqi77@126.com}
\affiliation{Shandong Key Laboratory of Space Environment and Exploration Technology, Institute of Space Sciences, School of Space Science and Technology, Shandong University, Shandong, China.}

\author{Dong-Yang Gao}
\email{gaodongyang@sdu.edu.cn}
\affiliation{Shandong Key Laboratory of Space Environment and Exploration Technology, Institute of Space Sciences, School of Space Science and Technology, Shandong University, Shandong, China.}

\author{Xiang Gao}
\email{202017698@mail.sdu.edu.cn}
\affiliation{Shandong Key Laboratory of Space Environment and Exploration Technology, Institute of Space Sciences, School of Space Science and Technology, Shandong University, Shandong, China.}

\author{Jing-Yi Wang}
\email{202217742@mail.sdu.edu.cn}
\affiliation{Shandong Key Laboratory of Space Environment and Exploration Technology, Institute of Space Sciences, School of Space Science and Technology, Shandong University, Shandong, China.}

\author{Ya-Ni Guo}
\email{yanierguo@163.com}
\affiliation{Shandong Key Laboratory of Space Environment and Exploration Technology, Institute of Space Sciences, School of Space Science and Technology, Shandong University, Shandong, China.}

\author{Xing Gao}
\email{34052688@qq.com}
\affil{Xinjiang Astronomical Observatory, 150 Science 1-Street, Urumqi 830011, China}

\author{Guo-You Sun}
\email{36723571@qq.com}
\affil{Xingming Observatory, Urumqi, Xinjiang, China}


\begin{abstract}
We present 22 sets of light curves and one radial velocity curve for the W UMa-type total eclipse contact binary system V2790~Ori, derived by combining all available public photometric data, the photometric data in previous studies, and our own spectroscopic and decade-long photometric observations. Our simultaneous analysis of the light curves and radial velocity curve shows that V2790 Ori is a W-subtype contact binary with a mass ratio of $q = 0.322(\pm0.001)$ and a shallow contact degree of $14.8(\pm0.6)\%$. The orbital period analysis based on 445 eclipsing minima reveals a secular decrease at a rate of $\dot P = -3.18 (\pm 0.75) \times 10^{-8}\mathrm{d~yr^{-1}}$, superimposed with a cyclic variation with an amplitude of $A = 8.98 (\pm 2.19) \times 10^{-4}~\mathrm{d}$ and a period of $ P_3 = 7.44 (\pm 0.52)~\mathrm{yr}$. The secular decrease is caused by AML via magnetic braking, while the cyclic period variation is explained by the light-travel time effect due to a third body, which is likely to be a brown dwarf. Furthermore, our analysis indicates a mass transfer from the more massive component to the less massive one at a rate of $1.22(\pm0.29) \times 10^{-8}~\mathrm{M_{\odot}~yr^{-1}}$. A model with a cool spot on each component was adopted to fit the O'Connell effect observed in the light curves. Since the O’Connell effect varies over time, we identified a solar-like magnetic activity cycle with a period of approximately 5.4~yr by analyzing the magnitude difference ($\Delta m$) at the two light maxima and the O’Connell effect ratio. In addition, evolutionary analysis suggests that V2790~Ori is a newly formed contact binary that evolved from a detached phase into the present contact configuration. 
\end{abstract}


\section{Introduction} 
Contact binaries are a class of close binary systems in which both components fill their Roche lobes \citep{1959cbs..book.....K}. W Ursae Majoris (W UMa)-type eclipsing binaries are a subset of contact binaries, characterized by both components sharing a common convective envelope (CCE; \citealt{1968ApJ...151.1123L,1968ApJ...153..877L}). Due to their contact configuration and tidal deformation, these systems exhibit continuous out-of-eclipse luminosity variations in their light curves. The presence of CCE enables efficient mass and energy exchange between the components, resulting in nearly equal surface temperatures despite different masses \citep{1968ApJ...151.1123L,1968ApJ...153..877L,1941ApJ....93..133K}. This leads to the primary and secondary eclipse depths of the light curve being nearly equal or differing insignificantly \citep{2014ApJS..212....4Q}. The light variations caused by eclipses impose strong constraints on the orbital and geometric properties of the system, making W UMa-type contact binaries ideal laboratories for determining the fundamental physical parameters of their components. This is crucial for investigating the physical characteristics of binary systems and testing theories of binary evolution \citep{2003MNRAS.342.1260Q, 2005ApJ...629.1055Y, 2007ApJ...662..596L, 10.1111/j.1365-2966.2008.13155.x, 2012JASS...29..145E}. W UMa-type contact binary systems are usually classified into two subtypes: A-subtype and W-subtype binaries \citep{1970VA.....12..217B}. Compared to A-subtype contact binaries, W-subtype contact binaries generally have shorter orbital periods \citep{1984QJRAS..25..405S}, later spectral types, and their less massive component tends to have a higher surface temperature.

The orbital periods of W UMa-type contact binary systems are generally shorter than 0.7 day \citep{2001icbs.book.....H}, and are not constant over time. The secular variation of orbital period is primarily governed by mass transfer and angular momentum loss (AML) via magnetic braking \citep{1971ARA&A...9..183P,1979A&A....80..287V, 1981A&A...102...81R, 1988MNRAS.231..823T}. Specifically, conservative mass transfer from the less massive to the more massive component results in an increase in the orbital period. In contrast, mass transfer in the reverse direction, AML, or a combination of AML and mass transfer can lead to a decrease in the period \citep[e.g.,][]{2020AJ....159..189L, 2024AJ....168..272P, 2025MNRAS.537.3160P}. In addition to secular trends, many W~UMa-type systems exhibit cyclic variations in their orbital period, as observed in V0599 Aur \citep{2020AJ....160...62H}, OO Leo \citep{2024ApJ...971..113M}, and LX~Lyn and V0853~Aur \citep{2024RAA....24a5022Z}. Such variations are often attributed to either magnetic activity cycle \citep{1992ApJ...385..621A} or the light-travel time effect (LTTE) caused by a third body \citep[e.g.,][]{1952ApJ...116..211I, 1959AJ.....64..149I, 1990BAICz..41..231M}. According to these theories, the Applegate mechanism proposes that magnetic activity cycle in one or both components leads to variations in the gravitational quadrupole moment, thereby modulating the orbital period. While the LTTE produces apparent period changes in orbital period as the eclipsing binary orbits the barycenter of a triple system.   

For most W UMa-type contact binary systems, the light curves exhibit asymmetries, which is known as the O’Connell effect \citep{1951PRCO....2...85O}. 
The O’Connell effect has been attributed to several mechanisms, such as starspots due to stellar magnetic activity \citep{1960AJ.....65..358B}, hot spots caused by a mass-transferring gas stream or material accretion between the components \citep{1994MmSAI..65...95S}, circumstellar material surrounding the binary \citep{2003ChJAA...3..142L}, and asymmetric circumfluence caused by Coriolis forces \citep{1990ApJ...355..271Z}. Among these, starspots are the most commonly used explanation for the O’Connell effect \citep{2009SASS...28..107W, 2014AJ....147...98L, 2022ApJS..262...10K}. Owing to their later spectral types, W-subtype contact binaries typically show more intense magnetic activity. In some contact binaries, the O’Connell effect remains nearly stable. However, similar to sunspots that follow an 11-year cycle, certain systems exhibit solar-like magnetic activity cycles, which can naturally lead to a quasi-periodic variation in the O’Connell effect \citep[e.g.,][]{2020PASJ...72...73L, 2023NewA..10102022Y, 2025RAA....25b5006W}. The long-term periodic variations in the orbital period, on timescales of several years to decades, may also serve as indirect evidence of such magnetic cycles. If W UMa-type contact binaries do possess magnetic activity cycles, long-term photometric observation is essential for their investigations.

V2790~Ori (TYC 1322-294-1, GSC 01322-00294) was first identified as a short-period W UMa-type contact binary system by \citet{2004IBVS.5570....1O} based on the public data release of the Northern Sky Variability Survey (NSVS; \citealt{2004yCat.2287....0W}). Its orbital period was reported to be $P = 0.287842~\mathrm{d}$ with a primary minimum at HJD~2451521.695 \citep{2004IBVS.5570....1O}. \citet{2006AJ....131..621G} estimated its distance to be 199~pc and reported that the $V$-band magnitude ranges from $V_{\mathrm{max}} = 11.181$ to $V_{\mathrm{min}} = 11.743$. V2790~Ori has been studied by \citet{2016JAVSO..44...30M}, \citet{2019RAA....19..143K}, \citet{2020NewA...8001400S}, and \citet{2021AcA....71..123A} based only on photometric observations spanning approximately one year. 
The determined physical parameters and orbital period variation analyses in previous studies are summarized in Table~\ref{t-previous}. However, because the photometric data and the eclipsing minima used for analysis cover different time spans, and the O’Connell effect in the light curves changes over time, their derived results are not identical. For example, while the derived physical parameters remain nearly unchanged, the derived spot parameters are not consistent. \citet{2019RAA....19..143K} reported an increase in the orbital period, whereas \citet{2021AcA....71..123A} found a period decrease superimposed with a cyclic variation. 
In addition, the significant and variable O’Connell effect in the light curves suggests the presence of a solar-like magnetic activity cycle.  The discrepancies among previous studies, together with the possible presence of a solar-like magnetic activity cycle, demonstrate the need for further detailed investigation of V2790~Ori. In this study, we combined all available photometric data with our own decade-long photometric and spectroscopic observations, constructing the most extensive data set for V2790 Ori to date. The simultaneous analysis of light and radial velocity curves enables us to derive reliable absolute physical parameters, providing a solid foundation for investigating the orbital period variation, magnetic activity cycle, and evolutionary status of V2790 Ori.

\section{OBSERVATIONS AND DATA REDUCTION} \label{sec:observe}
\subsection{Photometric Observations} \label{subsec:photo}
From 2015 to 2025, long-term multiband photometric observations of V2790~Ori were carried out by three ground-based telescopes located in China. The telescopes are: (1) the Weihai Observatory 1.0~m Cassegrain telescope of Shandong University (WHOT; \citealt{2014RAA....14..719H}), equipped with a PIXIS 2048B CCD, providing a field of view of 12$^\prime$×12$^\prime$; (2) the 60~cm Ningbo Bureau of Education and Xinjiang Observatory Telescope (NEXT), equipped with an FLI PL23042 CCD, producing a field of view of 22$^\prime$×22$^\prime$; (3) the 85~cm telescope at the Xinglong Station of the National Astronomical Observatories (XL85), equipped with an Andor DZ936 CCD, offering a field of view of 32$^\prime$×32$^\prime$. These telescopes are all equipped with Standard Johnson-Cousin-Bessel $BVR_{C}I_{C}$ filters. The detailed photometric observation log is listed in Table~\ref{t1}.

All CCD images were reduced using the PHOT package of IRAF\footnote[1]{IRAF is distributed by the National Optical Astronomy Observatory (NOAO), which is operated by the Association of the Universities for Research in Astronomy, Inc., under cooperative agreement with the National Science Foundation (NSF; \url{http://iraf.noao.edu/}.)}. After performing bias and flat corrections, a comparison star and a check star were selected in the same CCD field of view near V2790~Ori. The comparison star is 2MASS~J06154200+1938278 ($V_{\mathrm{mag}}=10.928$, $B-V = 0.724$, $J-K=0.481$ ), and the check star is 2MASS~J06151951+1937076 ($V_{\mathrm{mag}}=11.159$, $B-V = 0.144$, $J-K=0.185$ ). 
The differential magnitudes between the target and the comparison star, as well as between the comparison and check stars, were obtained using aperture photometry and differential photometry methods. 
Table~\ref{ta1} lists the Barycentric Julia Date (BJD) and the differential magnitudes (\(\Delta m\)) of the reduced observational data.

To more accurately investigate the variations of the O’Connell effect, we examined other available light curves. Photometric data from the American Association of Variable Star Observers (AAVSO)\footnote[2]{\url{https://www.aavso.org/}} and the Transiting Exoplanet Survey Satellite (TESS; \citealt{2015JATIS...1a4003R}) were analyzed in this study as well. We downloaded and organized all available data from the AAVSO International Database. The details of the light curves used for the subsequent analysis are listed in Table~\ref{t1}. Although \citet{2016JAVSO..44...30M} have analyzed the photometric data from AAVSO, their study only covers data from 2015. We also reanalyzed photometric data from the study of \citet{2019RAA....19..143K} and \citet{2020NewA...8001400S}.   

TESS observes approximately 150 million stars brighter than 16 magnitude in the TESS band, with a photometric precision ranging from 60 ppm to 3\%, enabling a wide array of stellar astrophysics investigations \citep{2018AJ....156..132O}. The photometric data are publicly available via the Mikulski Archive for Space Telescopes (MAST)\footnote[3]{\url{https://mast.stsci.edu/portal/Mashup/Clients/Mast/Portal.html}}. We searched the TESS data for V2790~Ori and found that it was observed in Sector 43, Sector 44, and Sector 45 at a 10-minute cadence, and in Sector 71 and Sector 72 at a 200-second cadence. Each sector was observed continuously for about 27.4 days, specifically spanning 2021 September 16 to October 11 (Sector 43), October 12 to November 5 (Sector 44), November 7 to December 2 (Sector 45), 2023 October 17 to November 11 (Sector 71), and 2023 November 12 to December 7 (Sector 72).  Additionally, the raw data in Sector 43, Sector 44, Sector 45, and Sector 71 have been calibrated by the Science Processing and Operations Center (SPOC) pipeline \citep{2016SPIE.9913E..3EJ}, which provides Pre-search Data Conditioning Simple Aperture Photometry flux (PDCSAP). However, the TESS-SPOC data in Sector 43 and Sector 45 are of relatively poor quality, and no TESS-SPOC data are available for Sector 71. Therefore, we carried out photometry for these sectors. We downloaded 20×20 pixel TESS full-frame image (FFI) cutouts \citep{2019ascl.soft05007B} from MAST using the lightkurve package \citep{2018ascl.soft12013L}. The aperture and background masks were then created to extract light curves by selecting an appropriate threshold, and the light curves were detrended using the Savitzky-Golay filter \citep{2007CSE.....9c..10O}. All the flux obtained were normalized and converted to magnitudes using the following equation: $mag = -2.5 \log(flux)$. To save computational time, the TESS data were binned to 1000 points by averaging. 

All light curves are phase folded using the following linear ephemeris:
\begin{equation}\label{eq1}
T = T_{0} + 0.287842 \times E,
\end{equation} 
where $T$ represents the observed time, $T_0$ is the primary minimum of each light curve, determined via the K-W method \citep{1956BAN....12..327K}, based on photometric data from WHOT, XL85, NEXT, AAVSO, and TESS . The orbital period of 0.287842~d is adopted from \citet{2004IBVS.5570....1O}, and $E$ is the cycle number. Finally, a total of 22 sets of phase-folded light curves are obtained. As shown in Figure~\ref{f-LC}, the primary minimum is distinctly flat-bottomed, indicating that V2790~Ori undergoes a total eclipse. The two maxima of each light curve are clearly unequal, and the difference between them varies over time. This phenomenon demonstrates that V2790~Ori exhibits a significant and time-variable O’Connell effect, which is worth further investigation. 

\subsection{Spectroscopic Observations} \label{subsec:spec}
A total of 20 spectra of V2790~Ori were obtained using the Beijing Faint Object Spectrograph and Camera (BFOSC) mounted on the 2.16 m telescope at Xinglong Observatory (XL216; \citealt{2016PASP..128k5005F}), including 8 spectra acquired on 2019 December 30 and 12 spectra on 2021 November 17. For BFOSC, the grating E9+G10 and a 1\farcs6-wide slit oriented in the south-north direction were employed. This setup provides a spectral resolution of $R \sim 11000$ per pixel, covering a wavelength range from 3300~\AA\ to 10000~\AA. We also obtained spectra of HD~65583 (K0, $V_{\mathrm{mag}}=7.00$), which serves as a radial velocity standard star. The exposure times of HD~65583 were set to 300 s for the 2019 observations and 200 s for the 2021 observations. Detailed information about the spectroscopic observations is listed in Table~\ref{t1}. 

The BFOSC spectroscopic data were reduced by the NOAO package in IRAF, including bias subtraction, flat correction, cosmic-ray removal, one-dimensional spectrum extraction, and wavelength calibration. Following these procedures, the normalized spectral data covering the wavelength range of 4706–5625~\AA\ were obtained. We used the cross-correlation function (CCF) \citep{2007A&A...465..943S, 2010AJ....140..184M} with HD~65583 to measure the radial velocities. The peak positions of the CCF curves were determined using the GaussPy+ package \citep{2019A&A...628A..78R}. A single peak is visible near phases 0 and 0.5, while double peaks appear at other phases. The derived radial velocities are listed in Table~\ref{ta2}. Phases were computed using Equation~(\ref{eq1}), adopting \( T_0 = \mathrm{BJD}~2458847.25856 \) for the 2019 observations and \( T_0 = \mathrm{BJD}~2459220.30069 \) for the 2021 observations. The radial velocity curve was then analyzed using rvfit\footnote[4]{\url{http://www.cefca.es/people/~riglesias/rvfit.html}} \citep{2015ascl.soft05020I}, an IDL-based code that fits radial velocity curves of binary systems through an adaptive simulated annealing global minimization algorithm. From the fitting, the mass ratio $q = 3.31 (\pm 0.08)$ and systemic velocity $V_\gamma = -43.84( \pm 0.56) \,\mathrm{km\,s^{-1}}$ were initially derived for V2790~Ori. 

\begin{deluxetable*}{ccccccc}
\tablecaption{Observation Log of V2790~Ori \label{t1}}
\tablewidth{0pt}
\tablehead{
\multicolumn{5}{c}{\text{Photometric Observations}} \\
\hline
\colhead{Date of Observation} & \colhead{Band (Exposure time (s))} & \colhead{Mean errors (mag)} & \colhead{Type} & \colhead{Data Source}
}
\startdata
2015 Feb 05 \& Mar 06           & $B(30)/V(20)/R_{C}(15)/I_{C}(10)$       & 0.007/0.008/0.007/0.009      & Light curve    & WHOT \\                                      
2015 Dec 07                     & $B(30)/V(15)$                           & 0.013/0.013                  & Minimum        & WHOT \\                                       
2015 Dec 28                     & $B(30)/V(15)/R_{C}(10)/I_{C}(8)$        & 0.010/0.008/0.006/0.008      & Light curve    & WHOT \\                                      
2016 Feb 20                     & $V(8)$                                  & 0.008                        & Minimum        & WHOT \\                                       
2016 Mar 30                     & $V(10)$                                 & 0.014                        & Minimum        & WHOT \\                                       
2016 Nov 05                     & $V(10)$                                 & 0.006                        & Minimum        & WHOT \\                                       
2016 Dec 30                     & $B(30)/V(15)/R_{C}(10)/I_{C}(8)$        & 0.010/0.020/0.007/0.007      & Light curve    & WHOT \\                                      
2017 Mar 01                     & $V(10)$                                 & 0.011                        & Minimum        & WHOT \\                                       
2017 Sep 11                     & $V(10)$                                 & 0.022                        & Minimum        & WHOT \\                                       
2017 Dec 20                     & $B(30)/V(15)/R_{C}(10)/I_{C}(8)$        & 0.012/0.007/0.006/0.007      & Light curve    & WHOT \\                                       
2017 Oct 01                     & $V(20)$                                 & 0.007                        & Minimum        & WHOT \\                                       
2018 Dec 27                     & $B(30)/V(20)/R_{C}(15)/I_{C}(15)$       & 0.007/0.005/0.006/0.006      & Light curve    & XL85 \\                                            
2019 Dec 29 \& 2020 Jan 03      & $B(18)/V(10)/R_{C}(7)/I_{C}(10)$        & 0.009/0.009/0.009/0.008      & Light curve    & NEXT \\               
2020 Dec 20                     & $B(30)/V(15)/R_{C}(8)/I_{C}(5)$         & 0.018/0.013/0.011/0.013      & Light curve    & WHOT \\ 
2021 Jan 04 \& Jan 05           & $B(18)/V(10)/R_{C}(7)/I_{C}(10)$        & 0.006/0.006/0.011/0.006      & Light curve    & NEXT \\            
2022 Jan 02                     & $B(30)/V(15)/R_{C}(10)/I_{C}(8)$        & 0.008/0.0011/0.008/0.010     & Light curve    & WHOT \\     
2022 Dec 31                     & $B(30)/V(12)/R_{C}(9)/I_{C}(8)$         & 0.004/0.0004/0.004/0.004     & Light curve    & XL85 \\  
2024 Feb 27 \& Mar 08           & $B(40)/V(20)/R_{C}(10)/I_{C}(8)$        & 0.006/0.006/0.007/0.004      & Light curve    & WHOT \\    
2025 Jan 08                     & $B(30)/V(12)/R_{C}(9)/I_{C}(8)$         & 0.007/0.0007/0.007/0.009     & Light curve    & XL85 \\ 
2015 Jan 08 -- Jan 29           & $g'/r'/i'$                              & ---                          & Light curve    & AAVSO (Edward Michaels) \\
2015 Nov 19 \& Nov 20           & $B/V$                                   & ---                          & Light curve    & AAVSO (Edward Michaels) \\
2016 Mar 12                     & $V$                                     & ---                          & Minimum        & AAVSO (Kenneth Menzies) \\
2017 Jan 02                     & $V$                                     & ---                          & Minimum        & AAVSO (Kenneth Menzies) \\
2020 Jan 29 -- Feb 05           & $B/V/I_{C}$                             & ---                          & Light curve    & AAVSO (Kevin Alton) \\    
2024 Sep 21                     & $V$                                     & ---                          & Minimum        & AAVSO (Sjoerd Dufoer) \\  
2015 Jan 21 -- Jan 23           & $B(120)/V(60)/R_{C}(60)$                & ---                          & Light curve    & TNO \\ 
2017 Jan 21 \& Jan 22           & $V/R_{C}/I_{C}$                         & 0.008/0.0008/0.009           & Light curve    & KAO \\
\hline
\multicolumn{5}{c}{\text{Spectroscopic Observations}} \\[2pt]
\hline      
Time (UT)                       & Number of Frames                     & Exposure (s)                 & Signal Noise Ratio    & Telescope \\[2pt]
\hline
2019 Dec 30                     & 8                                    & 900                          & 33 -- 43              & XL216 \\
2021 Nov 17                     & 12                                   & 900                          & 67 -- 92              & XL216 \\                                                               
\enddata
\tablecomments{
Mean errors for each band represent the photometric accuracy of the system, which were estimated by calculating the standard deviation of the differential magnitudes between the comparison and check stars; TNO data are from \citet{2019RAA....19..143K}; 
KAO data are from \citet{2020NewA...8001400S}.}
\end{deluxetable*}
                                                                                        
\section{RADIAL VELOCITY AND LIGHT CURVE ANALYSIS} \label{sec:WD}
In order to obtain the accurate physical parameters of V2790~Ori, we employed the 2015 version of the Wilson–Devinney (W-D) program \citep{1971ApJ...166..605W,1979ApJ...234.1054W,1990ApJ...356..613W}
to analyze the 22 sets of phase-folded light curves and radial velocity curve of V2790~Ori simultaneously. The initial value of the mass ratio was then set to $q = 3.31$, and the systemic velocity \( V_\gamma \) was initialized as \( -43.84 \,\mathrm{km\,s^{-1}} \).

It should be noted that both the rvfit and W-D programs adopt the mass ratio in the form of \( q = M_2/M_1 \) by default. Since the initial mass ratio is greater than 1, the less massive component is completely eclipsed by the more massive component during the total eclipse at phase 0. This enables a more precise determination of the effective temperature of the more massive component from the spectrum near phase 0. University of Lyon Spectroscopic analysis Software (ULySS; \citealt{2009A&A...501.1269K}) was employed to determine the atmospheric parameters of the more massive component. ULySS performs the spectral fitting by matching the observed spectrum to a model one generated by an interpolator based on the ELODIE stellar library \citep{2001A&A...369.1048P}. The atmospheric parameters are then derived by minimizing the residuals between the two spectra using a nonlinear least-squares algorithm. The observed spectrum obtained with BFOSC at phase 0.943 and the corresponding fitted spectra are shown in Figure~\ref{f-ULYSS} with black lines and red lines respectively. The obtained atmospheric parameters of the more massive component are as follows: $T_{\rm eff} = 5314 (\pm 42)\,\mathrm{K}$, \(\log g = 3.78 (\pm 0.09)\,\mathrm{cm\,s^{-2}}\), and $\mathrm{[Fe/H]} = -0.24 (\pm 0.04)\,\mathrm{dex}$. Finally, the effective temperature of the more massive component was fixed at 5314 K during W-D analysis, while that of the less massive component was adjustable. 

Because W UMa-type contact binaries share a CCE, we adopted gravity-darkening coefficients \( g_{1,2} = 0.32 \) \citep{1967ZA.....65...89L} and bolometric albedos \( A_{1,2} = 0.5 \) \citep{1969AcA....19..245R} for both components. The nonlinear limb-darkening law in the square-root form was adopted, and the bolometric and bandpass limb-darkening coefficients for both components were interpolated from the table provided by \citet{1993AJ....106.2096V}. Owing to the contact configuration of V2790~Ori, Mode 3 was selected during the modeling, which is appropriate for over-contact binaries where both component stars filling their Roche lobes. 
The following parameters are adjustable during the iterations: the orbital semi-major axis $a$, systemic radial velocity $V_\gamma$, orbital inclination $i$, mass ratio $q$, the effective temperature of the less massive component, the monochromatic luminosity of the less massive component, and the dimensionless potential of both components $\Omega_1 = \Omega_2$. The contact degree is calculated as \( f = (\Omega_\mathrm{in} - \Omega)/(\Omega_\mathrm{in} - \Omega_\mathrm{out}) \times 100\% \). 
Since the light curves of V2790~Ori exhibit an O’Connell effect, which is commonly attributed to magnetic activity of the components, the spot model was adopted in the W-D analysis. Different spot models were compared during the fitting process, and the optimal solutions were obtained by adding one cool spot on each component separately. The latitude of both spots was fixed at $90^{\circ}$, while other parameters including longitude, angular radius, and temperature factor were adjusted iteratively to reach the final converged solution. Meanwhile, we added the third light ($l_3$) in the W-D analysis in order to test the possible existence of a third companion. However, the solutions yielded a negative $l_3$ value, and no convergent solution was obtained.

\begin{figure*}[htb!]
\centering
\gridline{\fig{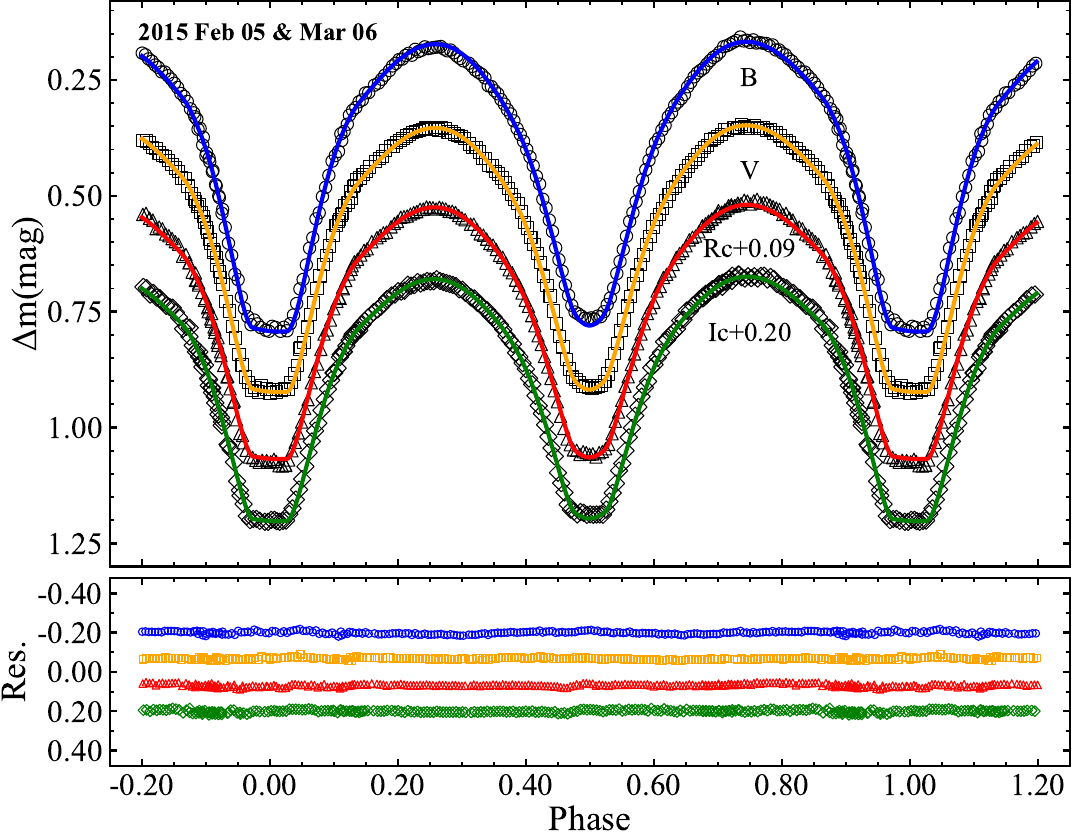}{0.30\textwidth}{(a)~2015 Feb 05 \& Mar 06 }
          \fig{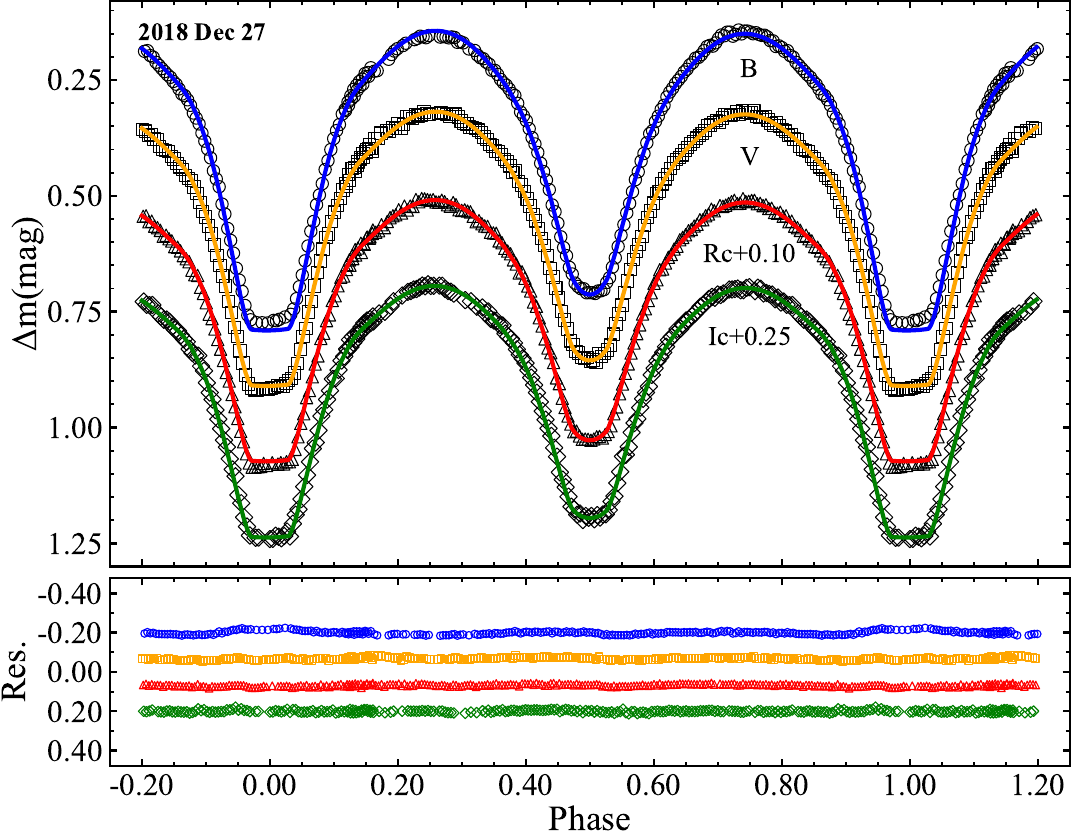}{0.30\textwidth}{(b)~2018 Dec 27}
          \fig{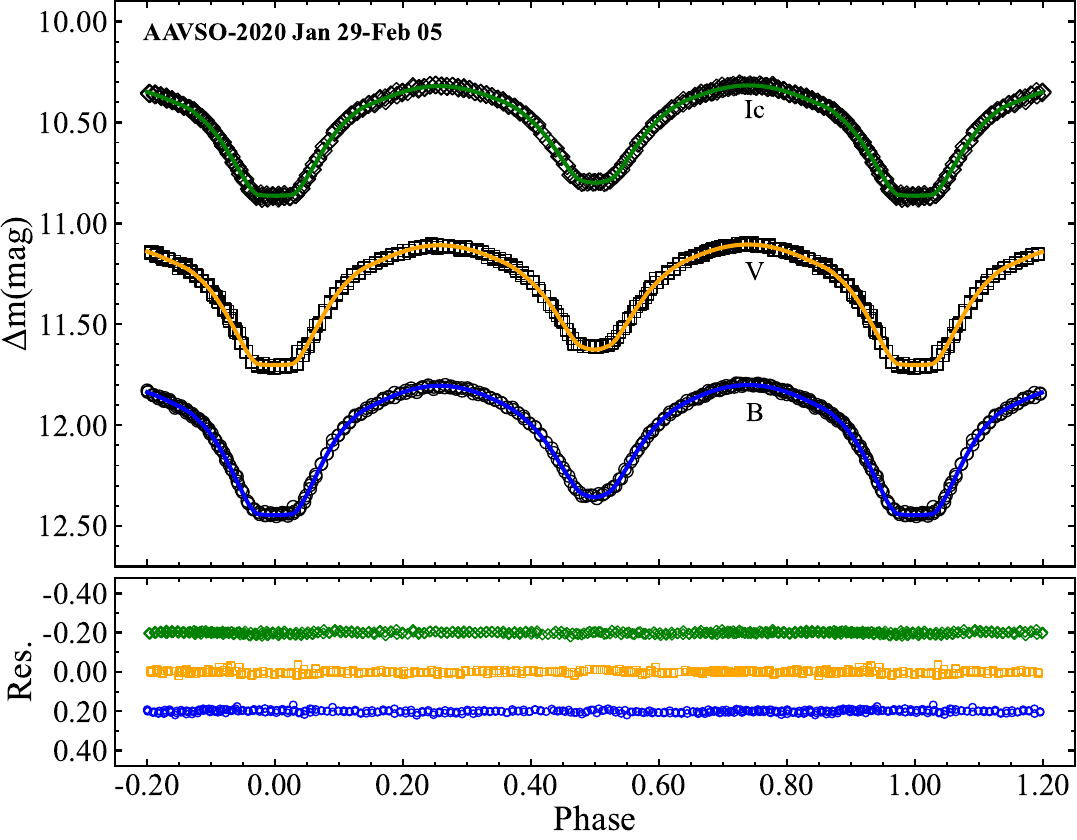}{0.30\textwidth}{(c)~AAVSO -- 2020 Jan 29 -- Feb 05}}
\vspace{-0.4cm}
\gridline{\fig{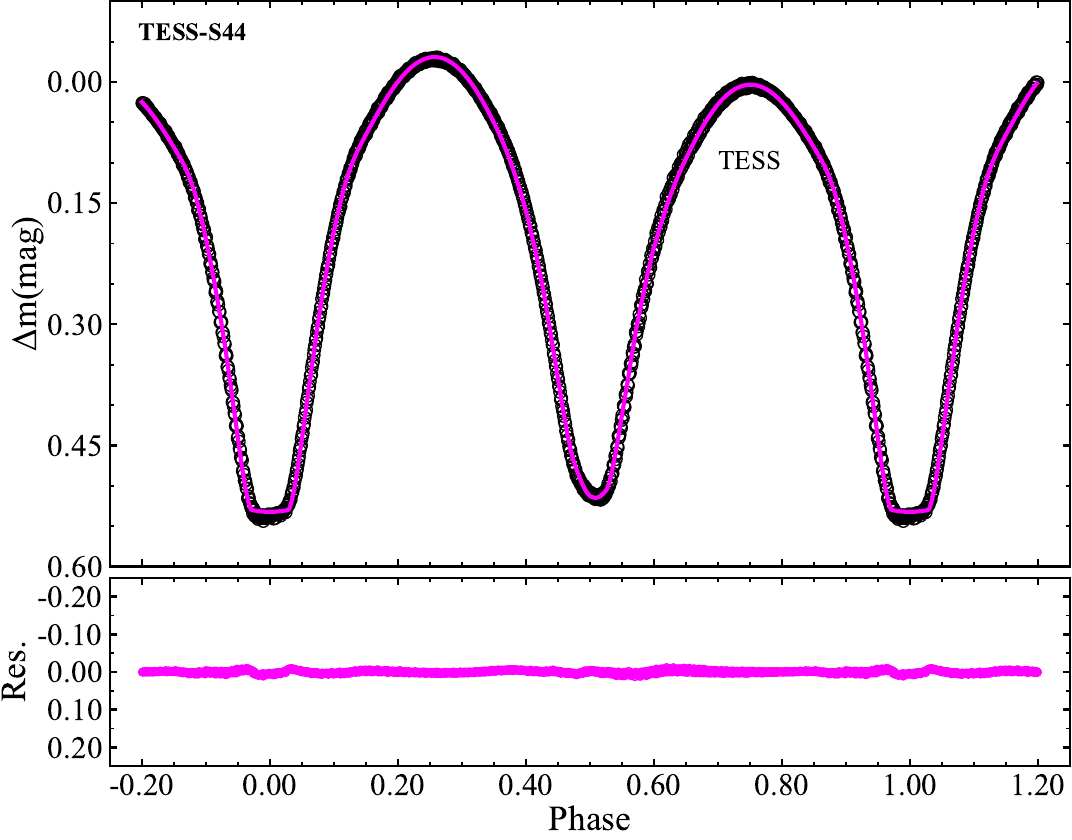}{0.30\textwidth}{(d)~TESS -- S44}
          \fig{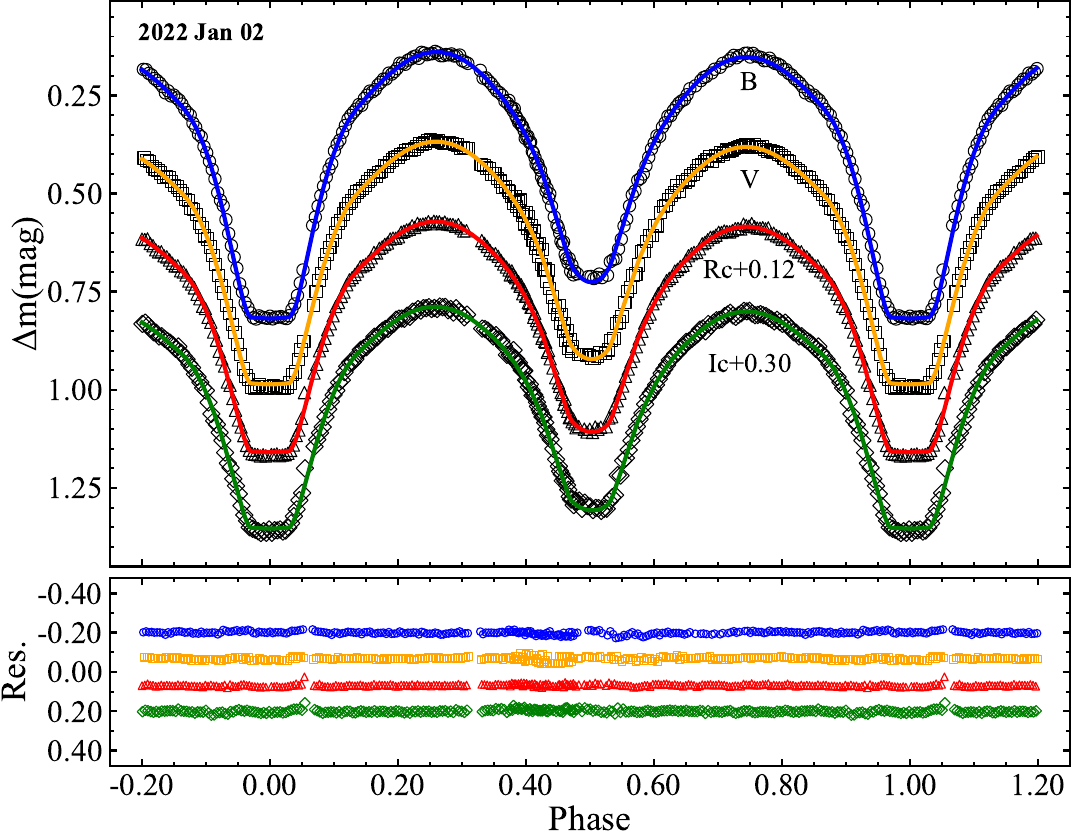}{0.30\textwidth}{(e)~2022 Jan 02}
          \fig{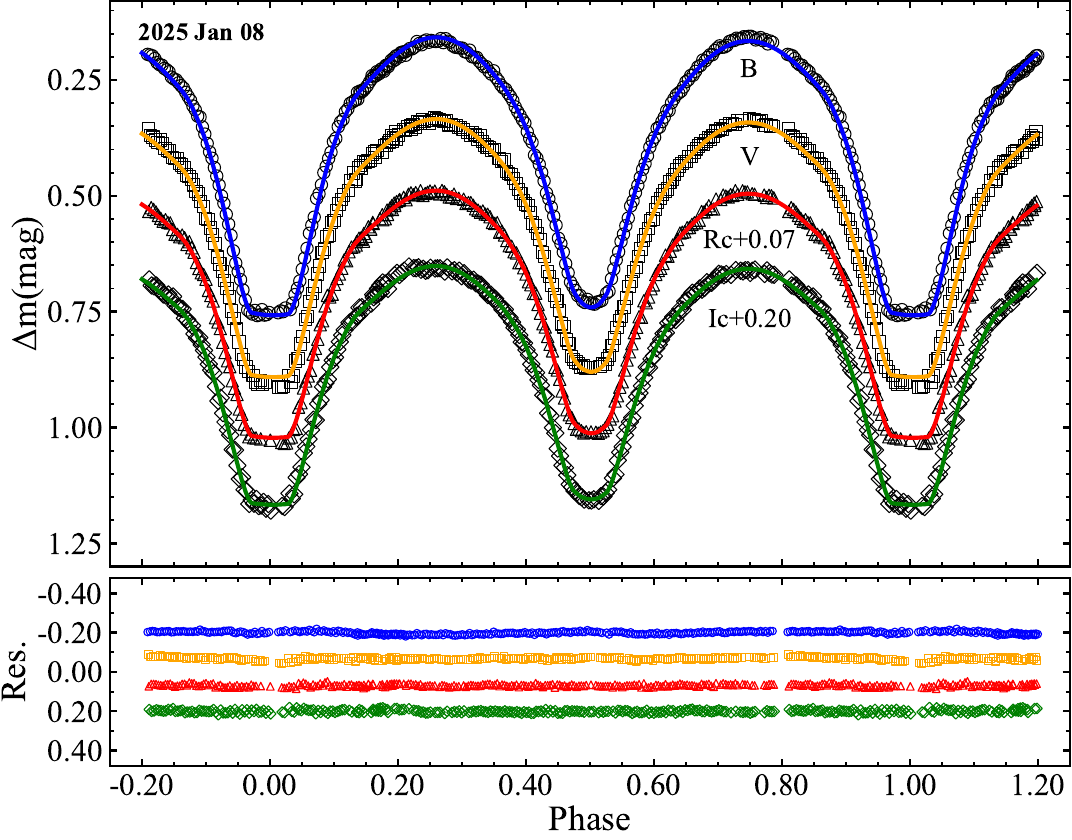}{0.30\textwidth}{(f)~2025 Jan 08}}
\caption{Theoretical light curves (solid lines) fitted by W-D program compared to observed ones for V2790~Ori. The fitted residuals are displayed at the bottom of each panel. All figures of 22 sets of light curves are available in the online journal.}
\label{f-LC}
\end{figure*}

\begin{figure*}[htb!]
\centering
\includegraphics[scale=0.47]{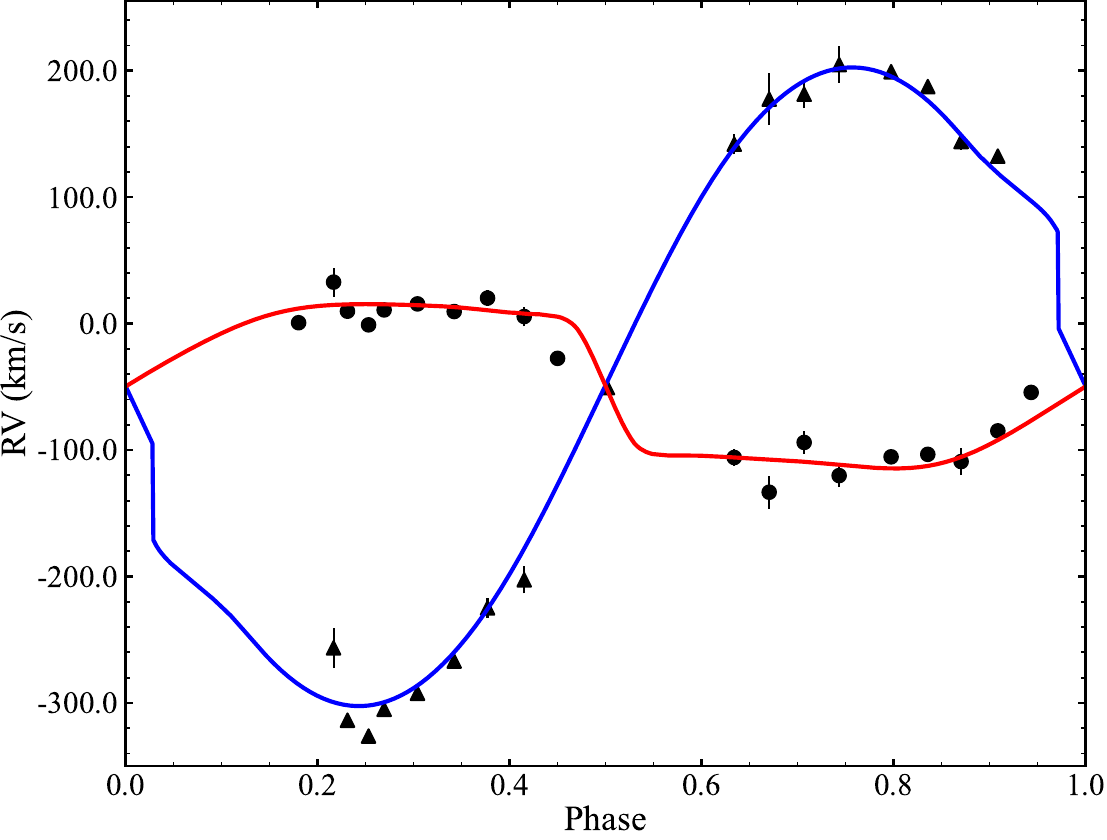}
\caption{Radial velocity and fitted curves of V2790~Ori. The black triangular represents the radial velocity of the less massive component, and the black circular represents the radial velocity of the more massive component. The blue lines show the theoretical curves of the less massive components, and the red lines show the theoretical curves of the more massive components.}
\label{f-RV}
\end{figure*}

For clarity and consistency in the subsequent analysis, we redefine the primary component (\( M_1\)) as the more massive star and the secondary component (\( M_2\)) is the less massive one, so that the adopted mass ratio is less than 1.
Finally, the orbital parameters derived from each set of light curves are summarized in Table~\ref{t-WD}, and the spot parameters are listed in Table~\ref{t-spot}. The uncertainties of the physical parameters represent the formal errors determined from the W-D program, which are underestimated \citep[e.g.,][]{1997PASP..109..782M, 2007A&A...467.1215S, 2020RAA....20...62A, 2021PASP..133h4202L}. We also analyzed all photometric data simultaneously using the W–D program, and the resulting photometric solution is listed in the "All data" column of Table~\ref{t-WD}, which is adopted as the final result. The theoretical and observed light curves and radial velocity curve are presented in Figure~\ref{f-LC} and Figure~\ref{f-RV}, respectively. 

\tabletypesize{\scriptsize}
\begin{deluxetable*}{lccccc}
\tablecaption{Photometric Solutions for Each Sets of Light Curves of V2790~Ori \label{t-WD}}
\tablewidth{0pt}
\tablehead{
\colhead{Parameters} & \colhead{2015 Feb 05 \& Mar 06} & 
\colhead{2015 Dec 28} & \colhead{2016 Dec 30} & 
\colhead{2017 Dec 20} & \colhead{2018 Dec 27}} 
\startdata
$a(R_\odot)$                              & 1.89(4)           & 1.90(4)       & 1.90(4)       & 1.89(4)      & 1.91(4)     \\
$V_\gamma(\text{km/s})$                   & -47.91(3)         & -47.99(3)     & -47.94(3)     & -47.81(3)    & -48.02(3)   \\
$q$                                       & 0.2923(11)        & 0.3082(11)    & 0.2908(11)    & 0.2784(5)    & 0.3165(10)  \\
$T_2(K)$                                  & 5530(6)           & 5468(8)       & 5418(7)       & 5495(5)      & 5591(4)     \\
$i(^{\circ})$                             & 83.0(1)           & 84.0(2)       & 82.7(2)       & 81.6(1)      & 84.8(2)     \\
$L_{2B}/(L_{1B}+L_{2B})$                  & 0.3019(12)        & 0.2949(20)    & 0.2763(17)    & 0.2808(13)   & 0.3327(9)   \\
$L_{2V}/(L_{1V}+L_{2V})$                  & 0.2884(9)         & 0.2853(15)    & 0.2700(13)    & 0.2699(10)   & 0.3146(7)   \\
$L_{2R_{C}}/(L_{1R_{C}}+L_{2R_{C}})$      & 0.2813(7)         & 0.2802(13)    & 0.2665(11)    & 0.2642(8)    & 0.3053(6)   \\
$L_{2I_{C}}/(L_{1I_{C}}+L_{2I_{C}})$      & 0.2763(6)         & 0.2765(11)    & 0.2641(10)    & 0.2601(7)    & 0.2987(6)   \\
$r_1$                                     & 0.4953(18)        & 0.4887(15)    & 0.4912(117)   & 0.4940(10)   & 0.4867(13)  \\
$r_2$                                     & 0.2831(5)         & 0.2853(4)     & 0.2832(4)     & 0.2738(3)    & 0.2888(3)   \\
$f(\%)$                                   & 11.9(2.9)         & 8.4(2.3)      & 9.2(2.6)      & 1.6(1.3)     & 10.5(1.9)   \\
\hline                        
Parameters            & 2019 Dec 29 \& 2020 Jan 03           & 2020 Dec 20        & 2021 Jan 04 \& Jan 05       & 2022 Jan 02       & 2022 Dec 31   \\       
\hline                                                                                                
$a(R_\odot)$                              & 1.91(4)          & 1.88(4)        & 1.92(4)       & 1.91(4)      & 1.89(4)     \\
$V_\gamma(\text{km/s})$                   & -48.03(3)        & -47.84(3)      & -48.04(3)     & -48.04(3)    & -47.96(3)   \\
$q$                                       & 0.3190(16)       & 0.2832(27)     & 0.3247(12)    & 0.3204(16)   & 0.3011(11)  \\
$T_2(K)$                                  & 5652(5)          & 5764(10)       & 5555(4)       & 5637(3)      & 5673(4)     \\
$i(^{\circ})$                             & 85.8(3)          & 83.2(3)        & 85.0(2)       & 86.3(2)      & 85.1(2)     \\
$L_{2B}/(L_{1B}+L_{2B})$                  & 0.3504(14)       & 0.3618(24)     & 0.3284(10)    & 0.3510(11)   & 0.3444(10)   \\
$L_{2V}/(L_{1V}+L_{2V})$                  & 0.3279(12)       & 0.3318(19)     & 0.3127(8)     & 0.3294(9)    & 0.3207(8)   \\
$L_{2R_{C}}/(L_{1R_{C}}+L_{2R_{C}})$      & 0.3165(11)       & 0.3171(18)     & 0.3045(8)     & 0.3185(9)    & 0.3089(7)   \\
$L_{2I_{C}}/(L_{1I_{C}}+L_{2I_{C}})$      & 0.3084(11)       & 0.3066(17)     & 0.2987(7)     & 0.3107(8)    & 0.3004(7)   \\
$r_1$                                     & 0.4879(24)       & 0.5102(46)     & 0.4856(17)    & 0.4930(15)   & 0.4918(16)  \\
$r_2$                                     & 0.2892(6)        & 0.2931(11)     & 0.2909(4)     & 0.2972(4)    & 0.2855(4)   \\
$f(\%)$                                   & 10.8(3.5)        & 32.4(6.8)      & 11.0(2.4)     & 22.9(2.2)    & 11.7(2.4)   \\       
\hline                        
Parameters            & 2024 Feb 27 \& Mar 08           & 2025 Jan 08        & 2015 Jan 08 -- 29       & 2015 Nov 19 \& 20       & 2020 Jan 29 -- Feb 05  \\       
\hline                                                                                                
$a(R_\odot)$                              & 1.91(4)          & 1.90(4)        & 1.90(4)       & 1.89(4)      & 1.90(4)     \\
$V_\gamma(\text{km/s})$                   & -48.01(3)        & -47.96(3)      & -47.94(3)     & -47.89(3)    & -47.92(3)   \\
$q$                                       & 0.3077(18)       & 0.3000(5)      & 0.2982(6)     & 0.2982(6)    & 0.2900(4)  \\
$T_2(K)$                                  & 5523(8)          & 5471(6)        & 5505(6)       & 5477(6)      & 5487(5)     \\
$i(^{\circ})$                             & 83.0(2)          & 83.8(1)        & 82.6(2)       & 82.2(2)      & 82.0(1)     \\
$L_{2B}/(L_{1B}+L_{2B})$                  & 0.3107(15)       & 0.2902(14)     & 0.2971(14)    & 0.2837(15)   & ---   \\
$L_{2V}/(L_{1V}+L_{2V})$                  & 0.2974(11)       & 0.2805(11)     & 0.2852(10)    & 0.2737(11)   & ---   \\
$L_{2R_{C}}/(L_{1R_{C}}+L_{2R_{C}})$      & 0.2904(10)       & 0.2753(9)      & ---           & ---          & ---   \\
$L_{2I_{C}}/(L_{1I_{C}}+L_{2I_{C}})$      & 0.2855(9)        & 0.2717(8)      & ---           & 0.2648(8)    & ---   \\    
$L_{2g'}/(L_{1g'}+L_{2g'})$               & ---              & ---            & ---           & ---          & 0.2856(10)   \\
$L_{2r'}/(L_{1r'}+L_{2r'})$               & ---              & ---            & ---           & ---          & 0.2709(6)   \\ 
$L_{2i'}/(L_{1i'}+L_{2i'})$               & ---              & ---            & ---           & ---          & 0.2751(7)   \\ 
$r_1$                                     & 0.4925(28)       & 0.4903(7)      & 0.4893(9)     & 0.4905(12)   & 0.4917(6)  \\
$r_2$                                     & 0.2887(9)        & 0.2823(3)      & 0.2805(4)     & 0.2766(4)    & 0.2810(2)   \\
$f(\%)$                                   & 15.0(4.3)        & 6.9(1.2)       & 4.1(1.5)      & 1.1(1.9)     & 7.2(0.9)   \\                 
\hline                        
Parameters            & TESS -- S43           & TESS -- S44        & TESS -- S45        & TESS -- S71       & TESS -- S72  \\       
\hline
$a(R_\odot)$                              & 1.93(4)          & 1.90(4)        & 1.91(4)       & 1.92(4)      & 1.91(4)     \\
$V_\gamma(\text{km/s})$                   & -48.06(3)        & -48.00(3)      & -48.03(3)     & -48.06(3)    & -48.05(3)   \\
$q$                                       & 0.3342(12)       & 0.3083(4)      & 0.3182(5)     & 0.3276(11)   & 0.3266(9)  \\
$T_2(K)$                                  & 5651(4)          & 5641(2)        & 5621(2)       & 5794(4)      & 5746(3)     \\
$i(^{\circ})$                             & 86.0(2)          & 84.5(1)        & 84.8(1)       & 88.3(2)      & 88.3(2)     \\
$L_{2T}/(L_{1T}+L_{2T})$                  & 0.3208(6)        & 0.3047(3)      & 0.3067(3)     & 0.3374(7)    & 0.3301(6)   \\
$r_1$                                     & 0.4876(16)       & 0.4950(5)      & 0.4905(7)     & 0.4899(14)   & 0.4908(13)  \\
$r_2$                                     & 0.2978(3)        & 0.2930(1)      & 0.2933(2)     & 0.2983(3)    & 0.2976(3)   \\
$f(\%)$                                   & 18.5(2.2)        & 21.3(0.7)      & 17.3(1.1)     & 21.5(2.0)    & 21.2(1.8)   \\      
\hline
Parameters            & TNO -- 2015 Jan 21 -- Jan 23           & KAO -- 2017 Jan 21 \& Jan 22        & All data        &       &   \\       
\hline
$a(R_\odot)$                              & 1.90(4)          & 1.93(4)        & 1.92(4)          &       &  \\
$V_\gamma(\text{km/s})$                   & -47.99(3)        & -48.06(3)      & -48.03(3)        &       &  \\
$q$                                       & 0.3056(7)        & 0.3313(20)     & 0.3221(3)        &       & \\
$T_2(K)$                                  & 5440(8)          & 5393(7)        & 5489(2)          &       &  \\
$i(^{\circ})$                             & 83.0(1)          & 85.5(3)        & 84.5(1)          &       &  \\
$L_{2B}/(L_{1B}+L_{2B})$                  & 0.2856(21)       & ---            & 0.3105(6)        &       &  \\
$L_{2V}/(L_{1V}+L_{2V})$                  & 0.2779(16)       & 0.2868(14)     & 0.2993(4)        &       &  \\
$L_{2R_{C}}/(L_{1R_{C}}+L_{2R_{C}})$      & 0.2737(13)       & 0.2840(12)     & 0.2934(4)        &       &  \\
$L_{2I_{C}}/(L_{1I_{C}}+L_{2I_{C}})$      & ---              & 0.2821(11)      & 0.2892(3)       &       &  \\  
$L_{2g'}/(L_{1g'}+L_{2g'})$               & ---              & ---             & 0.3054(5)       &       &  \\
$L_{2r'}/(L_{1r'}+L_{2r'})$               & ---              & ---             & 0.2943(4)       &       &  \\
$L_{2i'}/(L_{1i'}+L_{2i'})$               & ---              & ---             & 0.2899(3)       &       &  \\
$L_{2T}/(L_{1T}+L_{2T})$                  & ---              & ---             & 0.2898(3)       &       &  \\
$r_1$                                     & 0.4879(11)       & 0.4863(24)      & 0.4861(2)       &       & \\
$r_2$                                     & 0.2840(4)        & 0.2971(6)      & 0.2928(4)        &       &  \\
$f(\%)$                                   & 7.0(1.7)         & 18.3(3.3)       & 14.8(0.6)       &       & \\
\enddata
\end{deluxetable*}

The absolute physical parameters of V2790~Ori were determined directly through LC program as follows: the semi-major axis $a = 1.92(\pm0.04)~\mathrm{R_{\odot}}$, the masses of the two components $M_1 = 0.87(\pm0.01)~\mathrm{M_{\odot}}$ and $M_2 = 0.28(\pm0.01)~\mathrm{M_{\odot}}$, the radius $R_1 = 0.94(\pm0.02)~\mathrm{R_{\odot}}$ and $R_2 = 0.57(\pm0.01)~\mathrm{R_{\odot}}$, and the luminosity $L_1 = 0.63(\pm0.02)~\mathrm{L_{\odot}}$ and $L_2 = 0.26(\pm0.01)~\mathrm{L_{\odot}}$.


\section{ORBITAL PERIOD INVESTIGATION} \label{sec:ETV}
Investigating orbital period variations is an important approach to exploring the evolutionary status and potential companions of contact binaries \citep[e.g.,][]{2013NewA...21...46L, 2016Ap&SS.361...63L, 2019RAA....19..147L}. Such variations are typically analyzed using the $O\!-\!C$ method, which represents the difference between the observed and calculated eclipsing minima. 

\begin{figure*}[ht!]
\centering
\gridline{
  \fig{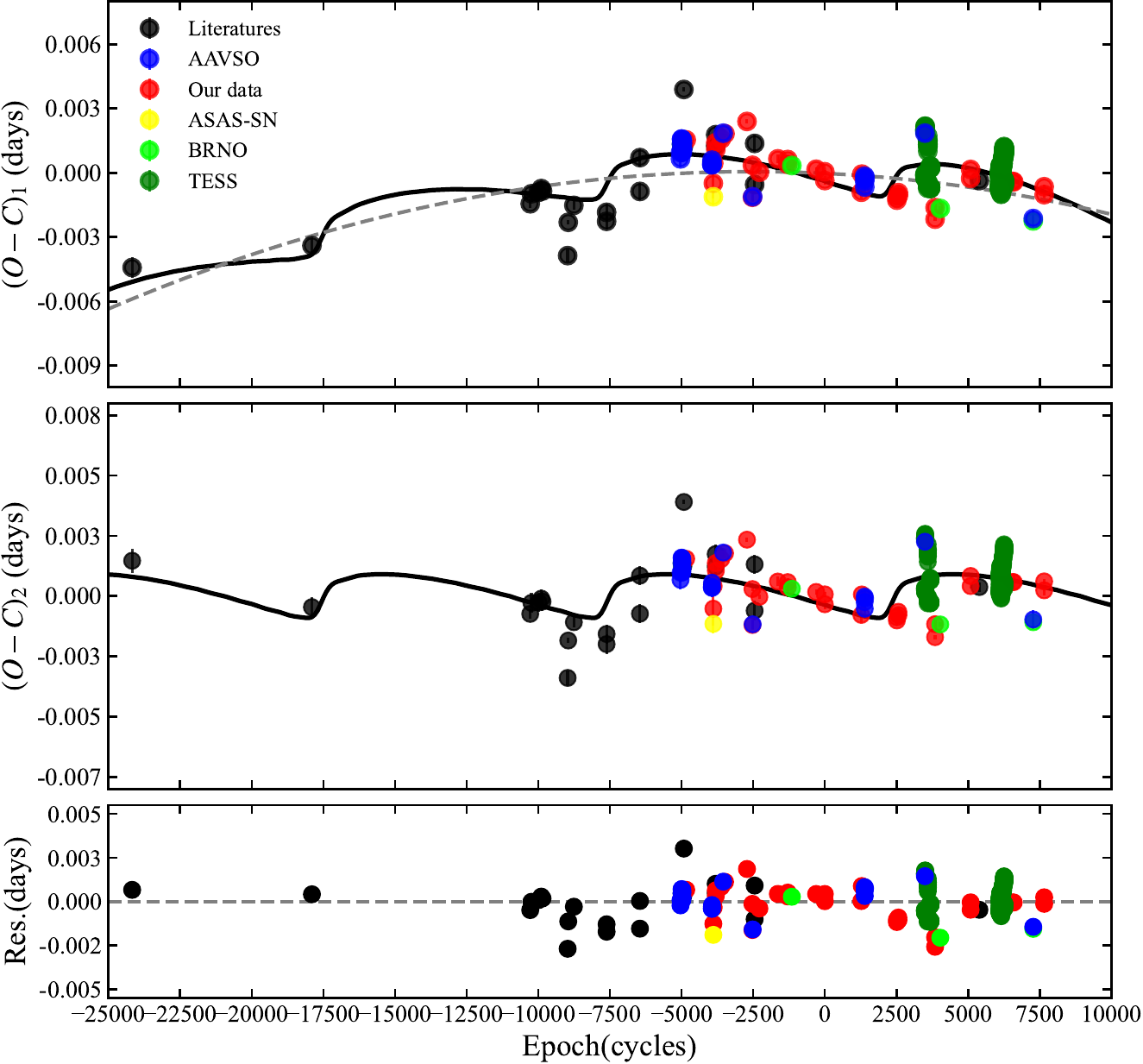}{0.49\textwidth}{(a)}
  \fig{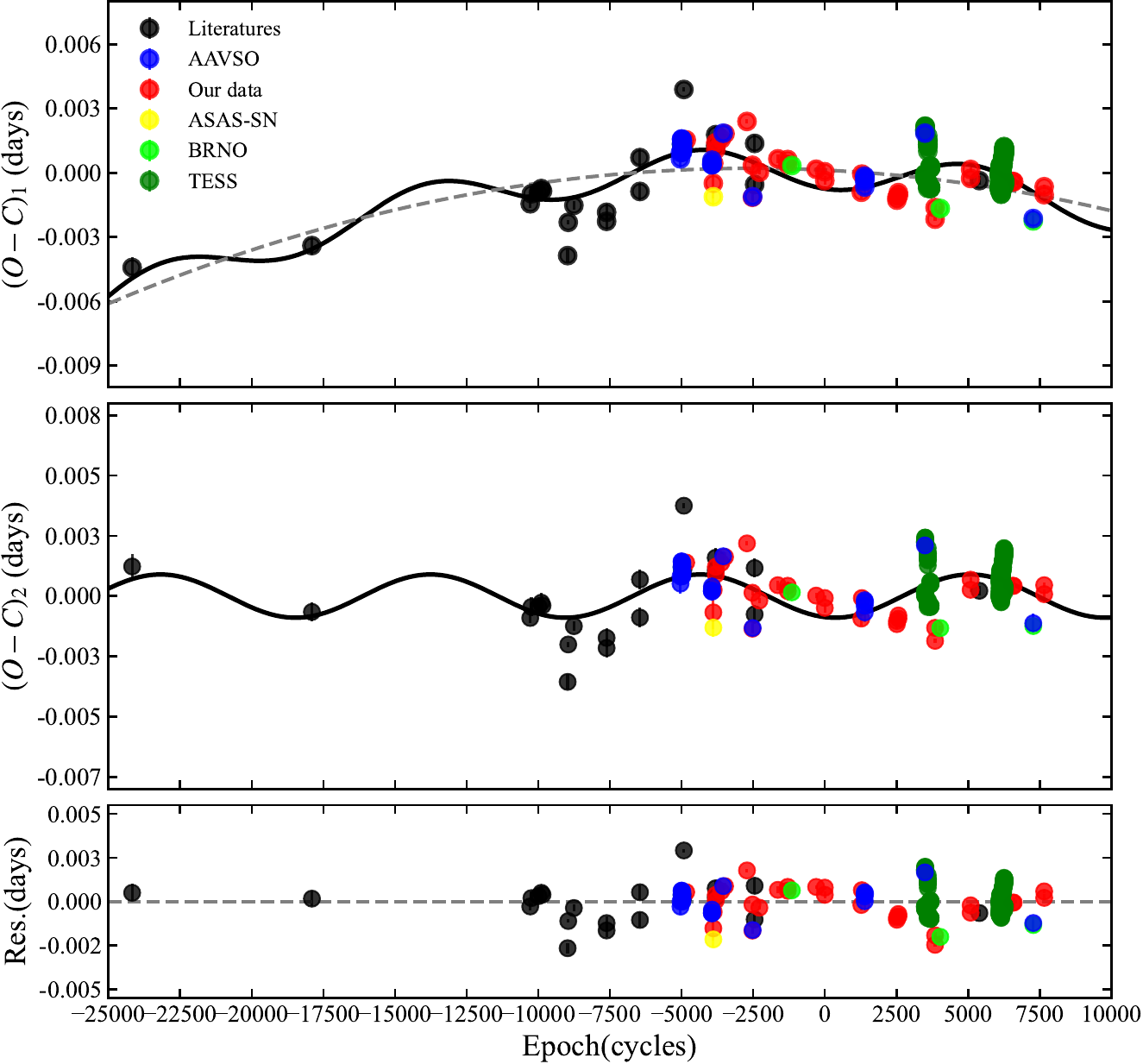}{0.49\textwidth}{(b)}
}
\caption{$O-C$ curves of V2790~Ori. (a)~Fitting result of Q+L model. (b)~Fitting result of Q+S model. The top panel shows the $(O-C)_1$ curve determined by the new linear ephemeris of Equation~(\ref{eq5}). The $(O-C)_2$ values, which remove the quadratic term from the $(O-C)_1$ curve, are plotted in the middle panel. The residuals from the full ephemeris of Equation (\ref{eq6}) or (\ref{eq7}) are displayed in the lower panel.}
\label{f-OC}
\end{figure*}

\citet{2004IBVS.5570....1O} first estimated the orbital period for V2790~Ori and derived the initial ephemeris as: $\mathrm{T}=2451521.695 (\mathrm{HJD}) + 0.287842\times E$. 
To construct a comprehensive $O\!-\!C$ diagram and further investigate the orbital period variations of V2790~Ori, we collected as many eclipsing minima as possible. In total, 445 minima spanning nearly 25 years were used in our analysis, including 31 from our photometric observations, 23 from AAVSO, 4 from Brno Regional Network of Observers project\footnote[5]{\url{http://var2.astro.cz/EN/brno/index.php}}(BRNO), 359 from TESS, 2 from the All-Sky Automated Survey for SuperNovae (ASAS-SN; \citealt{2014ApJ...788...48S, 2019MNRAS.486.1907J}), and 26 from other literatures. The K-W method \citep{1956BAN....12..327K} was used to calculate the eclipsing minima, except for those from other literatures. Since the observations of ASAS-SN and TESS at a 10-minute cadence are dispersed, we applied the phase shift method proposed by \citet{2020AJ....159..189L, 2021ApJ...922..122L, 2021AJ....162...13L} to compute eclipsing minima for these data. Because the minima of TESS data are given in Barycentric Julia Date (BJD), all other eclipsing minima were converted to BJD using the OSU Online Astronomy Utilities\footnote[6]{\url{http://astroutils.astronomy.ohio-state.edu/time/}}  provided by \citet{2010PASP..122..935E}. The following ephemeris was adopted to calculate the $O-C$ values:
\begin{equation} \label{eq4}
T = 2458480.26150\ (\mathrm{BJD}) + 0.287842\times E,
\end{equation}
where $T$ denotes the theoretical BJD corresponding to $E$ cycles. The initial epoch $\mathrm{BJD}~2458480.26150$ is from our observation on 2018 December 27. A linear least-squares fit to all minima yields the updated ephemeris:
\begin{equation} \label{eq5}
T = 2458480.26187(\pm0.00006)\ (\mathrm{BJD}) + 0.28784126(\pm0.00000002)\times E.
\end{equation}
Using this new ephemeris, we recalculated the $O\!-\!C$ values (denoted as $(O-C)_1$), which are listed in Table \ref{t-mini}. The corresponding $(O-C)_1$ curves are shown in the upper panel of Figure~\ref{f-OC}.

As shown in Figure~\ref{f-OC},  the $(O-C)_1$ curves exhibits a trend of periodic variation. This phenomenon can be interpreted either as the light-travel time effect (LTTE; \citealt{1922BAN.....1...93W, 1959AJ.....64..149I}) caused by a third body or as the magnetic activity cycle \citep{1992ApJ...385..621A}. We employed two models to fit the $(O-C)_1$ curves: (i) a quadratic-plus-sinusoid model (Q+S) and (ii) a quadratic-plus-LTTE model (Q+L), where LTTE corresponds to the five-parameter-LTTE solution described by \citet{1952ApJ...116..211I}. The two models are expressed as:
\begin{eqnarray}
(O-C)_1 &=& \Delta T_0 + \Delta P_0\times E + \frac{\beta}{2}\times E^2 + A \times \sin(\frac{2\pi}{P_3} \times E + \varphi),   \label{eq6} \\
\nonumber \\
(O-C)_1 & = & \Delta T_0 + \Delta P_0\times E + \frac{\beta}{2} \times E^2 + A \times \left[ \sqrt{1 - e^2} \sin E^* \cos \omega + \cos E^* \sin \omega \right],    \label{eq7}
\end{eqnarray}
Where $\Delta T_0$ and $\Delta P_0$ are the corrections of $T_0$ and $P_0$, respectively. The parameter $\beta$ denotes the secular change rate of the period, and $A$ and $P_3$ are the semi-amplitude and period of the periodic term. The parameters $e$, $\omega$, and $E^\ast$ denote the eccentricity, longitude of periastron, and eccentric anomaly of the third body's orbit, respectively \citep{1952ApJ...116..211I}. In fact, when the eccentricity $e = 0$, the LTTE model reduces to a sinusoidal form given in Equation~(\ref{eq6}). The Levenberg-Marquardt algorithm was employed to fit the $(O-C)_1$ curve of Q+L model. 
The final fitting curves are presented in Figure~\ref{f-OC}, and the fitted parameters from both models are listed in Table~\ref{t-ocfit}.

To determine the preferred model, we calculated the residual sum of squares (RSS) for both models and employed the Akaike Information Criterion (AIC) and Bayes Information Criterion (BIC). 
The results of RSS, AIC, and BIC are summarized in Table~\ref{t-ocfit}. By comparing these values, the RSS, AIC and BIC of Q+S model is smaller, suggesting that it offers a superior fit to the data. Based on this model, the orbital period is decreasing at a rate of $\dot P = -3.18 (\pm 0.75) \times10^{-8}\mathrm{d~yr^{-1}}$, accompanied by a periodic variation with an amplitude of $A = 8.98 (\pm 2.19) \times 10^{-4}~\mathrm{days}$ and a period of $P_3 = 7.44 (\pm 0.52)~\mathrm{yr}$. 

\tabletypesize{\small}
\begin{deluxetable*}{ccc}
\tablecaption{Fitting Results of Two Models for $O-C$
\label{t-ocfit}}
\tablewidth{0pt}
\tablehead{
\colhead{Parameter} & \colhead{Q+S} & \colhead{Q+L}
}
\startdata
$\Delta T_0$ (days)         & $1.25(\pm1.80)\times10^{-4}$      & $-2.65(\pm26.33)\times10^{-5}$   \\        
$\Delta P_0$                & $-6.44(\pm2.34)\times10^{-8}$     & $-6.42(\pm1.22)\times10^{-8}$     \\       
$\dot P$ (d\,yr$^{-1}$)     & $-3.18(\pm0.75)\times10^{-8}$     & $-3.22(\pm0.32)\times10^{-8}$     \\       
$A$ (days)                  & $8.98(\pm2.19)\times10^{-4}$      & $1.74(\pm4.42)\times10^{-3}$     \\        
$e$                         & ---                               & $0.888(\pm0.611)$                   \\     
$\omega$ ($^{\circ}$)       & ---                               & $6.00(\pm0.64)$                     \\     
$P_3$ (yr)                  & $7.44(\pm0.52)$                   & $8.05(\pm1.26)$                    \\      
$T_3$                       & ---                               & $2453517(\pm10561)$             \\         
$\phi$ ($^{\circ}$)         & $256.26(\pm11.01)$                & ---                               \\       
RSS                         & 0.000245                          & 0.000248                         \\        
AIC                        & -6401.5                            & -6392.1                             \\     
BIC                        & -6376.9                            & -6359.3                             \\     
\enddata
\end{deluxetable*}

\section{DISCUSSIONS AND CONCLUSIONS} \label{sec:result}
From the spectroscopic and decade-long multiband photometric observations of V2790~Ori, we derived its physical parameters through a simultaneous analysis of the light and radial velocity curves using the W-D program. 
The criteria for classifying contact binaries as deep, medium, or shallow are as follows: $f\geq 50\%$ is deep, $25\% \leq f < 50\%$ is medium, and $f< 25\%$ is shallow \citep{2005AJ....130..224Q, 2022AJ....164..202L, 2023MNRAS.519.5760L}. Therefore, V2790~Ori is a W-subtype shallow contact binary with a mass ratio of $q = 0.322(\pm0.001)$ and a contact degree of $14.8(\pm0.6)\%$, which is generally consistent with the results of previous studies as shown in Table~\ref{t-previous}. Based on the effective temperatures of both components, the spectral type of the primary component is estimated to be G9 and that of the secondary G7 \citep{2013ApJS..208....9P}. 

\subsection{Orbital Period Variations and Their Physical Origins}  \label{subsec:period}
\citet{2019RAA....19..143K} and \citet{2021AcA....71..123A} have previously investigated the orbital period variations of V2790~Ori, as summarized in Table~\ref{t-previous}. \citet{2019RAA....19..143K} reported an orbital period increase based on 36 eclipsing minima spanning nearly 12 years, attributing it to mass transfer from the less massive component to the more massive component. Earlier, \citet{2016JAVSO..44...30M} had detected a small cyclic change but lacked enough data to perform the necessary calculations. More recently, \citet{2021AcA....71..123A} found a period decrease superimposed with a cyclic variation based on eclipsing minima covering about 20 years, which was interpreted as the LTTE induced by a third body. In this study, we collected 445 eclipsing minima spanning nearly 25 years. Our analysis shows that the orbital period of V2790~Ori is undergoing a secular decrease at a rate of $\dot{P} = -3.18 (\pm 0.75) \times 10^{-8}~\mathrm{d~yr^{-1}}$, superimposed with a cyclic variation with an amplitude of $A = 8.98 (\pm 2.19) \times 10^{-4}~\mathrm{d}$ and a period of $P_3 = 7.44 (\pm 0.52)$~yr. 

\subsubsection{Secular Period Decrease} \label{subsubsec:secular}
In general, the continuous orbital period decrease can be attributed to AML and mass transfer from the more massive component to the less massive one. Stellar wind-driven magnetic braking can substantially contribute to the AML process. 
Based on the magnetic braking model \citep{1967ApJ...148..217W}, \citet{1988ASIC..241..345G} derived the orbital period decrease rate as follows:
\begin{equation}  \label{eq10}
\dot{P}_{\mathrm{aml}} \approx -1.1 \times 10^{-8} \times \frac{(1+q)^{2}}{q}\times \frac{k_{1}^{2}
M_{1}R_{1}^{4}+k_{2}^{2}M_{2}R_{2}^{4}}{(M_{1}+M_{2})^{5/3}P^{7/3}},
\end{equation}
where $k_{1}$ and $k_{2}$ are the gyration radius of the primary and secondary components, respectively. During the calculations, $k_{1}$ was determined using the relations derived by \citet{2009A&A...494..209L}: $k_{1}=0.5391 - 0.2504 \times M_{1}~(0.4M_{\odot}<M_{1}<1.4M_{\odot})$, where both the rotation and tidal distortions of the two components were taken into account. Because the mass of the secondary component is very low, it should be a fully convective star, $k_{2}^2= 0.205$ was used \citep{2007MNRAS.377.1635A}. By substituting the physical parameters into Equation (\ref{eq10}), the period decrease rate due to AML was calculated to be $\dot{P}_{\mathrm{aml}} = -6.57 \times 10^{-8}~\mathrm{d~yr^{-1}}$. This value is close to the observed rate of $\dot P = -3.18 (\pm 0.75) \times10^{-8}\mathrm{d~yr^{-1}}$, suggesting that the AML is sufficient to account for the secular decrease of orbital period. Assuming the conservation of mass and angular momentum, the mass transfer rate can be estimated using the following relation \citep{1975MNRAS.170..633P}:
\begin{equation}  \label{eq11}
\dot{M_1} = -\dot{M_2} = \frac{M_1M_2}{3P(M_1-M_2)} \times \dot{P}.
\end{equation}
Accordingly, the mass transfer rate from the primary to the secondary component is calculated to be $\dot{M_1} = -1.22(\pm0.29) \times 10^{-8}~\mathrm{M_{\odot}~yr^{-1}}$.

\subsubsection{Cyclic Period Variation} \label{subsubsec:cyclic}
As shown in the middle panel of Figure~\ref{f-OC}, the $(O-C)_{2}$ curve exhibits a clear cyclic variation. Such cyclic variation observed in the orbital period of eclipsing binaries are commonly explained by the LTTE caused by a third body \citep[e.g.,][]{2014AJ....147...98L,  2018PASP..130g4201L, 10.1093/mnras/staf1170,10.1093/mnras/staf216} or the magnetic activity cycle of one or both components \citep{1992ApJ...385..621A}. According to statistical analysis, W~UMa-type contact binaries have a high probability to belong to multiple systems \citep{2010MNRAS.405.1930L, 1992IAUS..151..315C}. Therefore, assuming that the cyclic variation in the orbital period of V2790~Ori is caused by a third body, the projected distance between the eclipsing binary and the barycenter of the triple system is given by $a_{12} \sin i_3 = A \times c$, where $a_{12}$ is the orbital radius of the binary around the center of mass of the triple system, $i_3$ is the inclination of the third body’s orbit, and $c$ is the speed of light. By inserting the amplitude and period of the cyclic variation, $a_{12}sin{i_3}$ is computed to be $ 0.16(\pm0.04)~\mathrm{AU}$. Using the following equation:  
\begin{equation} \label{eq12}
f(M_{3}) = \frac{4\pi}{GP^2_{3}}\times (a_{12}\sin i_{3})^{3} = 
\frac{(M_{3}\sin i_{3})^{3}}{(M_{1}+M_{2}+M_{3})^{2}},
\end{equation}
the mass function of the third body was calculated as $f(M_3) = 6.80(\pm4.98) \times 10^{-4}~\mathrm{M_{\odot}}$. The orbital distance between the third body and the inner binary can be estimated using $ a_3 = (M_1+M_2) \times a_{12}/M_3$. Assuming a coplanar orbit ($i_3 = i = 84^{\circ}.5$), the mass of the third body is determined to be $M_3 = 0.046 \pm 0.019~\mathrm{M_{\odot}}$, and the corresponding orbital distance is $a_3 = 4.57 \pm 2.21~\mathrm{AU}$. This suggests that the third component is a brown dwarf, whose upper mass limit is approximately $0.072~\mathrm{M_{\odot}}$ \citep{Basri2023}. 

Applegate’s theory \citep{1992ApJ...385..621A} proposes that magnetic activity cycle in a binary system can drive cyclic variation in the orbital period by inducing internal structural variations, which are coupled to the orbit via tidal interactions. Both the photometric solutions and the variable O’Connell effect are indicative of the magnetic activity in V2790~Ori. If the magnetic activity mimics the the variation in the orbital period and results in the observed cyclic variation, the required variation of the gravitational quadrupole moment $\Delta Q$ can be determined using the following equation \citep{2002AN....323..424L}:
\begin{equation} \label{eq13}
\frac{\Delta P}{P} = \frac{2\pi \times A}{P_3} = -9 \frac{\Delta Q}{Ma^2},
\end{equation}
where $\Delta P$ is the period of cyclic variation, $P$ is the orbital period, $M$, $R$, and $a$ denote the mass, radius of the magnetically active components, and the semi-major axis of the binary system, respectively. Based on this relation, we derived $\Delta Q_1 = 7.07\times 10^{48} ~\mathrm{g~cm^2}$ and $\Delta Q_2 = 2.30\times 10^{48} ~\mathrm{g~cm^2}$ for the primary and secondary components. However, under the assumption of angular momentum conservation, the typical value of $\Delta Q$ is on the order of $10^{51}$-$10^{52}~\mathrm{g~cm^2}$ for close binaries and $10^{49}~\mathrm{g~cm^2}$ for the cataclysmic variables \citep{1999A&A...349..887L}. The values of $\Delta Q_1$ and $\Delta Q_2$ obtained for V2790~Ori are significantly lower than these typical ranges, suggesting that the Applegate mechanism cannot adequately explain the cyclic period variation observed in this system. Therefore, such variation can be attributed to the LTTE caused by a third body, which is a brown dwarf. It is worth noting that no third light has been detected in either our analysis or previous studies of V2790~Ori \citep{2016JAVSO..44...30M, 2021AcA....71..123A}, or it may be too faint to be identified through LC modeling, which is consistent with the presence of a brown dwarf companion.

\subsection{O’Connell effect variation} \label{subsec:oconnell}
Given the interpretation of the O’Connell effect as being caused by active spots, it may serve as a reliable indicator of magnetic activity. Theoretically, the period of magnetic activity cycle should be consistent with the variation period of the O’Connell effect. In the case of V2790~Ori, the different spot solutions from our (Table~\ref{t-spot}) and previous studies (Table~\ref{t-previous}), together with the 22 sets of light curves presented in Figure~\ref{f-LC} that exhibit variable O’Connell effect, enable the investigation of a potential magnetic activity cycle.

In order to quantify the O’Connell effect, we first calculated the magnitude difference between the primary and secondary maxima ($\Delta m$) in each light curve, which is a conventional metric. In addition to this basic measurement, we also employed the O’Connell effect ratio (OER), as proposed by \citet{1997ASPC..130..129M}. OER is defined as the ratio of the areas beneath the two maxima of a phase-folded light curve. For the numerical analysis, the phase-folded light curve is initially divided into $n$ equal-width phase bins, with $n$ chosen to ensure adequate sampling of the light curve. The average magnitude within each bin is then computed and normalized by subtracting the mean magnitude of the bin corresponding to the primary minimum. The OER is finally calculated using the following equation:
\begin{equation}
OER = \frac{\sum_{k=1}^{n/2} (m_k - m_1)}{\sum_{k=n/2+1}^{n} (m_k - m_1)},
\end{equation}
where $m_k$ denotes the mean magnitude in the $k$th bin. Compared to the conventional $\Delta m$ measurement, the OER provides a more sensitive diagnostic for detecting asymmetries in the out-of-eclipse regions of a light curve. Notably, it has been proven to be more valuable for tracing the continuous variation of light-curve asymmetry over time \citep{2009SASS...28..107W}, and has been adopted in numerous studies \citep[e.g.,][]{ 2020AJ....160...62H, 2021AAS...23741807K, 2024ApJ...971..113M}. 

\begin{figure*}[htb!]
\centering
\includegraphics[scale=0.4]{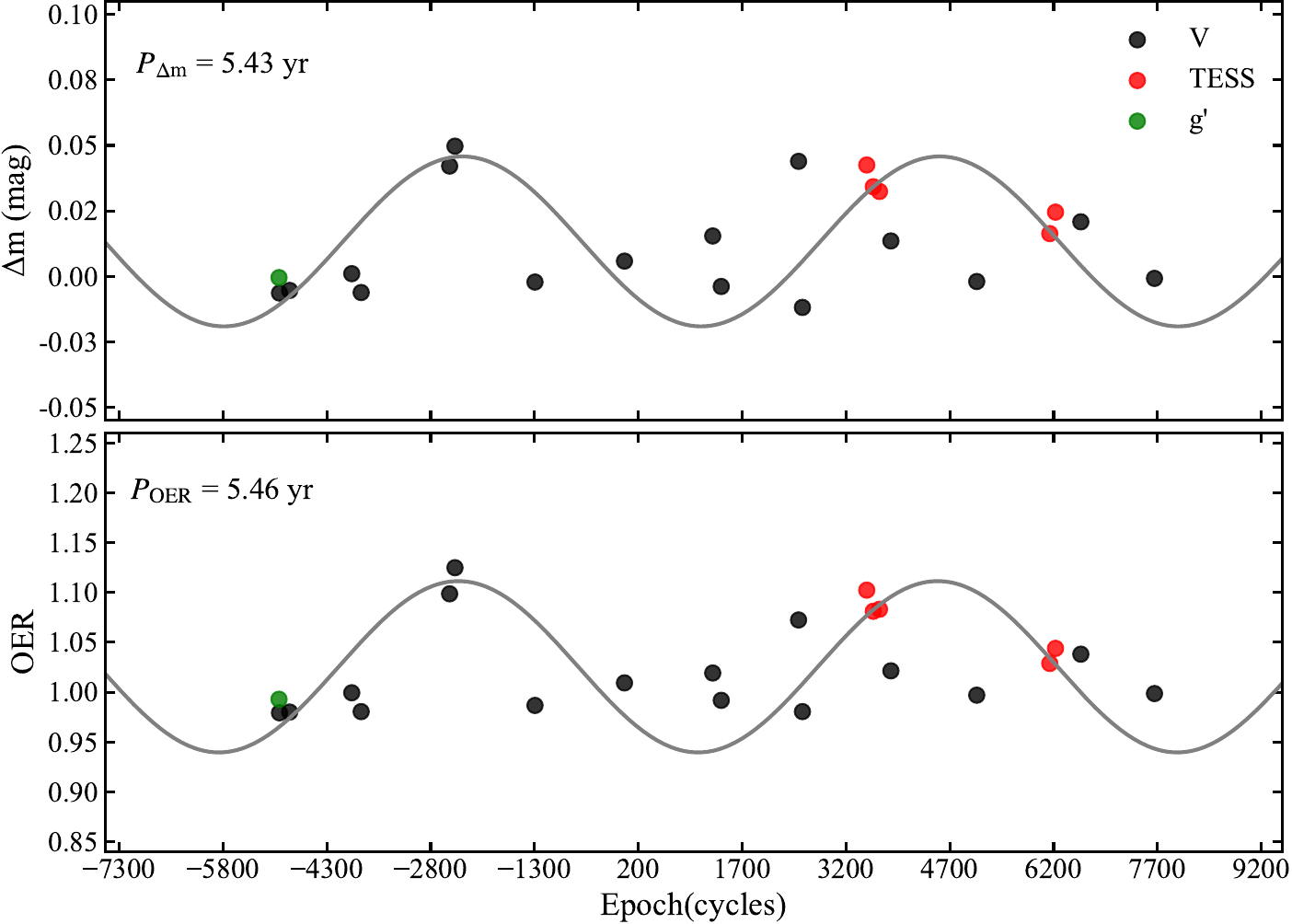}
\caption{Possible cyclic variations for $\Delta m$ and OER. The solid lines represent the sinusoidal fit with Equation (\ref{16}) and Equation (\ref{17}). The periods determined are presented in the top left corner of each panel.}
\label{f-OER}
\end{figure*}

Using the 22 available light curves, we calculated both $\Delta m$ and OER values of all bands, with results summarized in Table~\ref{t-OER}. To ensure consistency in investigating the long-term variation of the O’Connell, we selected the V band for most light curves, the $g^\prime$ band for one AAVSO light curve, and the TESS band for 5 light curves from TESS. To search for the cyclic variation in the O’Connell effect, we fitted both $\Delta m$ and OER data using the following sinusoidal function:   
\begin{equation}  \label{15}
f(x) = a + A_m \times \sin\left( \frac{2\pi}{P_{\mathrm{m}}} \times E + \varphi \right),
\end{equation}
where $A_m$, $P_{\mathrm{m}}$, and $\varphi$ represent the amplitude, period and initial phase of the sine term, respectively. $E$ denotes the cycle number, consistent with our previous definition. Through least-squares fitting, we obtained the following final expressions:
\begin{equation} \label{16}
\Delta m = 0.01332(\pm 0.01137) + 0.03245(\pm 0.01502) \times \sin\left( \frac{2\pi}{5.43(\pm 0.68)} \times E - 212.28(\pm 26.65) \right),
\end{equation} 

\begin{equation} \label{17}
OER = 1.02547(\pm 0.00272) + 0.08589(\pm 0.03611) \times \sin\left( \frac{2\pi}{5.46(\pm 0.61)} \times E - 214.75(\pm 23.67) \right),
\end{equation}
and the fitting curves are presented in Figure~\ref{f-OER}. The derived periods of variation in $\Delta m$ and OER are $5.43(\pm 0.68)~\mathrm{yr}$ and $5.46(\pm 0.61)~\mathrm{yr}$, respectively. This indicates that V2790~Ori likely exhibits a solar-like magnetic activity cycle with a period of approximately 5.4 years. The period of the observed cyclic period variation ($ P_3 = 7.44 (\pm 0.52)~\mathrm{yr}$) differs from that of the magnetic activity cycle, which further confirms that the magnetic activity cycle is insufficient to explain the cyclic period variation. To further confirm this magnetic activity cycle, more observations in the future are required for verification.

\subsection{Evolutionary State} \label{subsec:evo}
\begin{figure*}[ht!]
\centering
\gridline{
  \fig{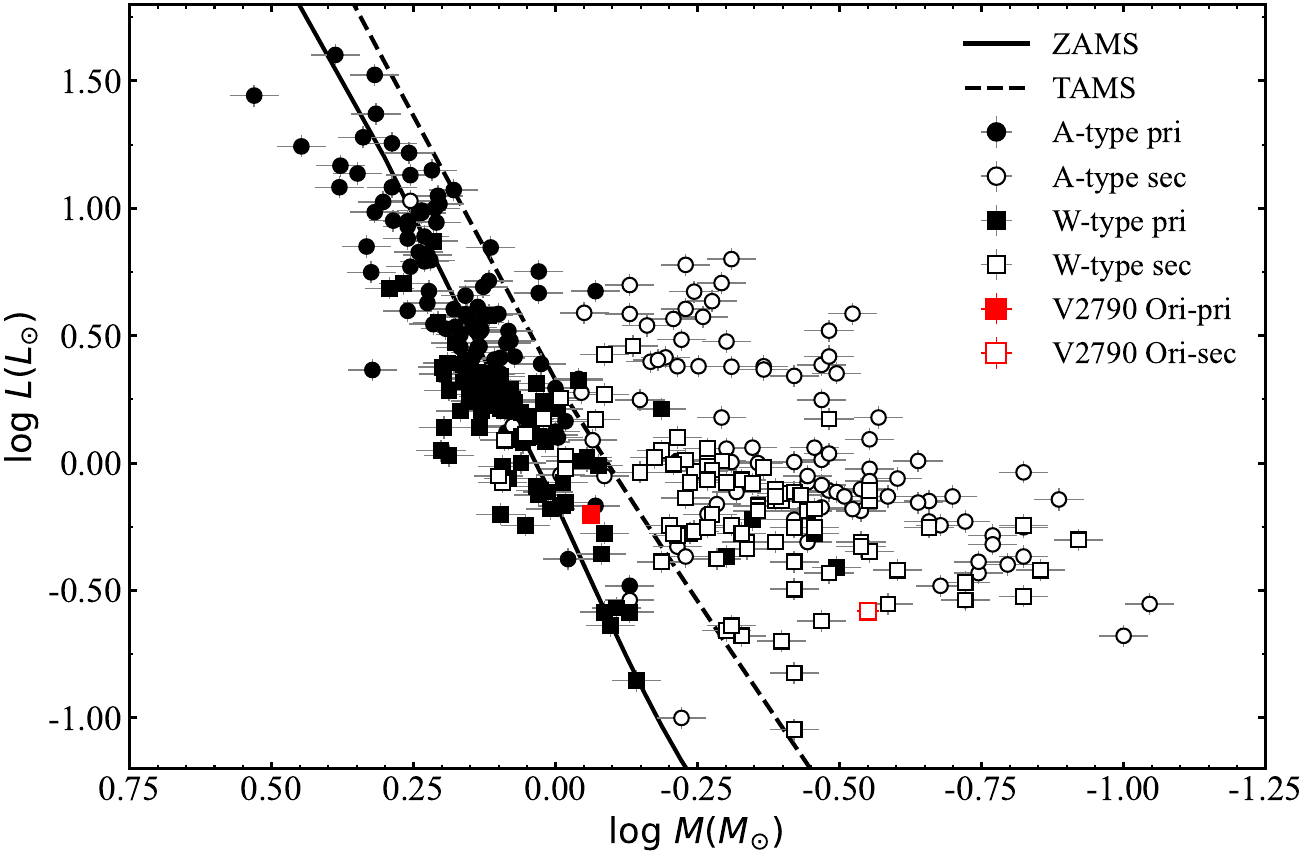}{0.45\textwidth}{(a)}
  \fig{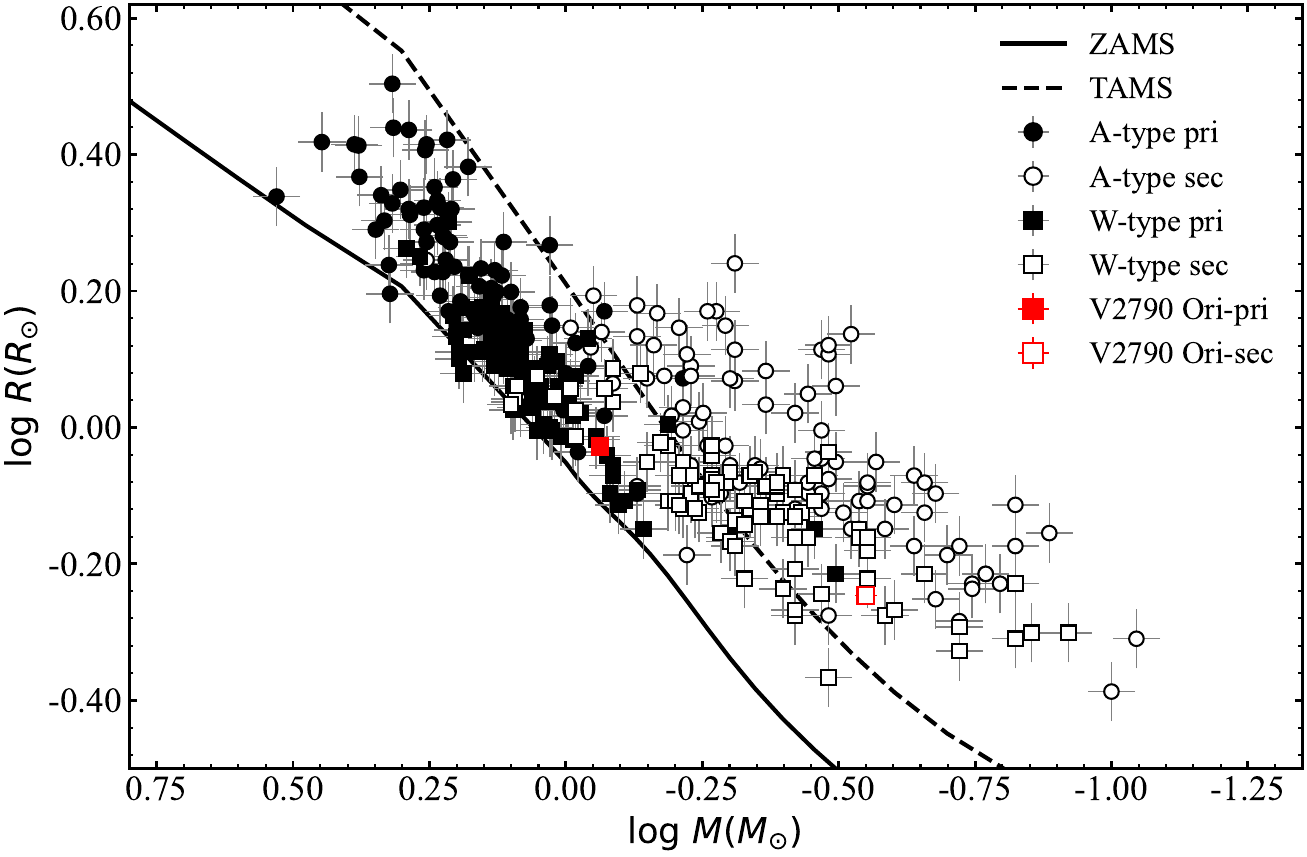}{0.45\textwidth}{(b)}}
\vspace{-0.4cm}
\gridline{\fig{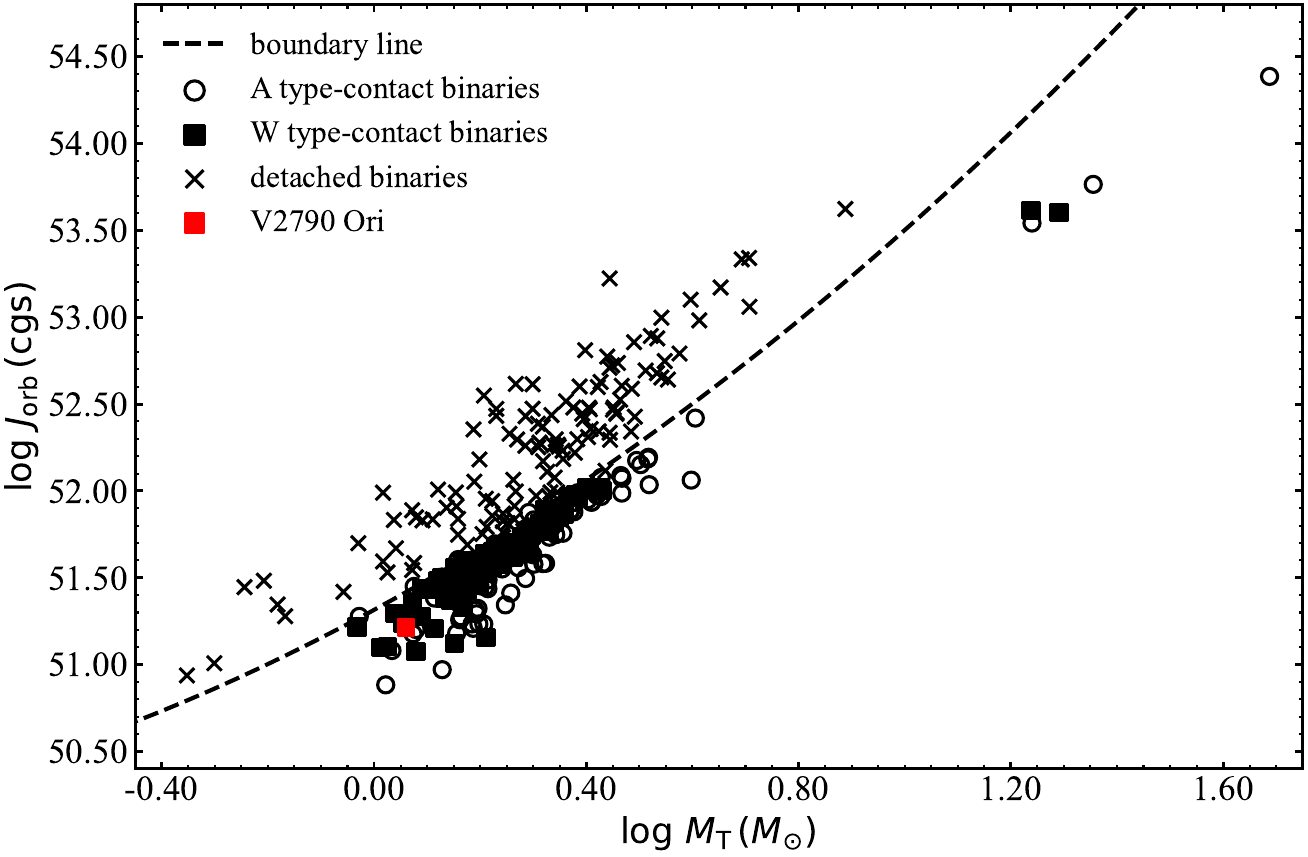}{0.45\textwidth}{(c)}}
\caption{(a)~The relation of M-L. (b)~The relation of M-R. The zero-age main sequence (ZAMS) and terminal-age main sequence (TAMS), generated using the binary star evolution code from \citet{2002MNRAS.329..897H}, are plotted as solid and dashed black lines, respectively. Circular and square symbols represent W-subtype and A-subtype contact binaries, respectively, and are taken from \citet{2021AJ....162...13L}. Solid symbols denote the more massive primary components, while open symbols represent the less massive secondary components. The positions of the components of V2790~Ori are marked with red squares.
(c)~The relation between orbital angular momentum and total mass for detached and contact binaries. The dashed boundary line between detached and contact binaries and detached binaries are adopted from \citet{2006MNRAS.373.1483E}. The samples of contact binaries are taken from \citet{2021AJ....162...13L}. Open circular and solid square symbols represent A-subtype and W-subtype contact binaries, respectively. The position of V2790~Ori is marked with a red solid square.}
\label{f-MR}
\end{figure*}

In order to further investigate the evolutionary status of both components, the mass-radius and the mass-luminosity diagrams were constructed based on the absolute physical parameters of V2790~Ori as shown in Figure~\ref{f-MR}. 
Although no reliable evolutionary model has been established for contact systems, the zero-age and terminal-age main sequence (ZAMS and TAMS) lines adopted in our analysis were derived from the binary star evolution code of \citet{2002MNRAS.329..897H}. As shown in Figure~\ref{f-MR}, V2790~Ori lies within the distribution region of W-subtype contact binaries, where the primary component is located between the ZAMS and TAMS lines, whereas the secondary component is above the TAMS line.
This implies that the primary component is still on the main sequence, while the secondary component has evolved off the main sequence and appears oversized and over-luminous compared to a main-sequence star of the same mass. This may be attributed to energy and mass transfer from the primary to the secondary component during the early evolutionary stages of the system.

The orbital angular momentum $J_{orb}$ of V2790~Ori was calculated using the following equation from \citet{2013AJ....146..157C}:
\begin{equation} \label{eq18}
J_{\mathrm{orb}} = 1.24 \times 10^{52} \times M_{\mathrm{T}}^{3/5} \times P^{1/3} \times q \times (1 + q)^{-2},
\end{equation}
where $q$ is the mass ratio, $M_{\mathrm{T}} = M_{1} + M_{2}$ is the total mass of the binary system, and $P$ is the orbital period. The orbital angular momentum of V2790~Ori is computed to be $\log J_{\mathrm{orb}} = 51.2~\mathrm{cgs}$. Then the diagram of $M_{\mathrm{T}}$ versus $J_{\mathrm{orb}}$ is shown in Figure~\ref{f-MR}, and the position of V2790~Ori is located below and near the boundary line. In combination with the shallow contact degree of V2790~Ori, this suggests that the system is a newly formed contact binary. Moreover, its current orbital angular momentum is lower than that of detached systems with the same total mass, supporting the interpretation that angular momentum and mass loss during the earlier detached phase led the binary to evolve into contact configuration.

The progenitors of both components of V2790~Ori were estimated using the following equations from \citet{2013MNRAS.430.2029Y}: 
\begin{equation} \label{eq19}
M_{\mathrm{Si}} = M_{\mathrm{S}} + \Delta M,
\end{equation}

\begin{equation} \label{eq20}
M_{\mathrm{Pi}} = M_{\mathrm{P}} - (\Delta M - M_{\mathrm{lost}}) = M_{\mathrm{P}} - \Delta M (1 - \gamma),
\end{equation}

\begin{equation} \label{eq21}
\Delta M = 2.50 \times \left[ \left( \frac{L_{\mathrm{S}}}{1.49} \right)^{1/4.216} - M_{\mathrm{S}} - 0.07 \right]^{0.64},
\end{equation}    
where $M_{\mathrm{Pi}}$ and $M_{\mathrm{Si}}$ are the initial masses, and $M_{\mathrm{P}}$ and $M_{\mathrm{S}}$ are the current masses of the primary and secondary components, respectively. $L_{\mathrm{S}}$ is the luminosity of the secondary component. $\Delta M$ is defined as the total mass lost by the secondary component, $M_{\mathrm{lost}}$ refers to the mass lost from the system, and $\gamma$ is the ratio $M_{\mathrm{lost}}/\Delta M$. Using the observed values of $M_{\mathrm{S}} = 0.282 \pm 0.013~\mathrm{M_{\odot}}$ and $L_{\mathrm{S}} = 0.262 \pm 0.012~\mathrm{L_{\odot}}$, the initial mass of the secondary component is calculated as $1.465 \pm 0.048~\mathrm{M_{\odot}}$, and the total mass lost by the secondary is $1.183 \pm 0.035~\mathrm{M_{\odot}}$. \citet{2013MNRAS.430.2029Y} suggested that the value of $\gamma$ for W-subtype contact binaries lies within the range 0.500-0.664, and Figure~\ref{f-MM} shows the relation between present total mass $M_{\mathrm{T}}$ and initial total mass $M_{\mathrm{Ti}}$ for A-subtype and W-subtype contact binaries from \citet{2014MNRAS.437..185Y}. The positions of V2790~Ori corresponding to the two boundary $\gamma$ values are marked with a red solid hexagon and a red solid triangle, respectively. As shown in Figure~\ref{f-MM}, the case where $\gamma= 0.664$ lies closer to the fitted line for W-subtype systems. Thus, the appropriate value of $\gamma$ for V2790 Ori should be 0.664.
So the initial mass of the primary component is determined as $M_{\mathrm{Pi}} = 0.468(\pm 0.008)~\mathrm{M_{\odot}}$. Initially, V2790~Ori was in a detached configuration. The more massive progenitor of the current secondary component evolved more rapidly toward TAMS and filled its Roche lobe first, which led to the beginning of mass transfer. The progenitor of the secondary component transferred $0.397 \pm 0.012~\mathrm{M_{\odot}}$ to the progenitor of the primary component, resulting in a reversal of the mass ratio. Meanwhile, a total of $0.785(\pm 0.024)~\mathrm{M_{\odot}}$ has been lost from the system since the detached phase.  Eventually, the system evolved into the current W-subtype contact binary system. Additionally, \citet{2021AcA....71..123A} suggested that the initial orbital period of V2790~Ori lies in the range of 2-4 days,  corresponding to the peak of the orbital period distribution predicted by the influence of Kozai cycles \citep{1962AJ.....67..591K} coupled with tidal friction \citep{2007ApJ...669.1298F}, under the assumption that the system contains a third body \citep{2006Ap&SS.304...75E,2007AJ....134.2353R}. This further proves that the observed cyclic variation in the orbital period is caused by a third body.

In summary, we conducted spectroscopic and decade-long photometric observations of V2790~Ori. The physical parameters of the system were determined through a simultaneous analysis of its light curves and radial velocity curve. The orbital period was investigated using 445 eclipsing times, revealing a secular decrease superimposed with a cyclic variation.The secular decrease is attributed to AML, while the periodic variation, with a period of $ P_3 = 7.44 (\pm 0.52)~\mathrm{years}$, is explained by the LTTE induced by third body, which is most likely to be a brown dwarf. Based on the long-term variation of the O’Connell effect, a solar-like magnetic activity cycle with a period of approximately $5.4~\mathrm{years}$ was identified through the analysis of $\Delta m$ and OER using all available light curves. Evolutionary analysis suggests that V2790~Ori is a newly formed W-subtype contact binary system, which has evolved from a detached phase into the present contact configuration via a combination of mass transfer, mass loss, and AML. Therefore, in order to robustly confirm the magnetic activity cycle and orbital period variations of V2790~Ori, continuous photometric observations will be essential in the future.  


\begin{acknowledgments}

We are grateful to the anonymous referee for the constructive comments and valuable suggestions that significantly improved this manuscript. This work is supported by the National Natural Science Foundation of China (NSFC, No.~12273018), the Joint Research Fund in Astronomy (No.~U1931103) under a cooperative agreement between NSFC and the Chinese Academy of Sciences (CAS), the Taishan Scholars Young Expert Program of Shandong Province, the Qilu Young Researcher Project of Shandong University, and the Young Data Scientist Project of the National Astronomical Data Center. The calculations in this work were carried out at Supercomputing Center of Shandong University, Weihai.

We acknowledge the support of the staff of the Xinglong 85~cm, 2.16~m telescopes, NEXT and WHOT. This work was partially supported by National Astronomical Observatories, Chinese Academy of Sciences. 

This work includes data collected by the TESS mission. Funding for the TESS mission is provided by NASA Science Mission Directorate. The specific observations analyzed in this work can be accessed via the MAST at DOI: \href{https://doi.org/10.17909/fwdt-2x66}{10.17909/fwdt-2x66}. We acknowledge the TESS team for its support of this work. We gratefully acknowledge the variable star observations from the AAVSO International Database, contributed by observers worldwide and used in this research. We also appreciate the observers around the world for providing the BRNO data of V2790~Ori. We thank Las Cumbres Observatory and its staff for their continued support of ASAS-SN. ASAS-SN is funded in part by the Gordon and Betty Moore Foundation through grants GBMF5490 and GBMF10501 to the Ohio State University, and also funded in part by the Alfred P. Sloan Foundation grant G2021-14192. This work has made use of data from the European Space Agency (ESA) mission Gaia (https://www.cosmos.esa.int/gaia), processed by the Gaia Data Processing and Analysis Consortium (DPAC; https://www.cosmos.esa.int/web/gaia/dpac/consortium). Funding for the DPAC has been provided by national institutions, in particular the institutions participating in the Gaia Multilateral Agreement. 
\end{acknowledgments}

\appendix

\setcounter{table}{0}        
\renewcommand{\thetable}{A\arabic{table}} 
\setcounter{figure}{0}        
\renewcommand{\thefigure}{A\arabic{figure}} 

\begin{figure*} [htb!]
\centering
\includegraphics[scale=0.5]{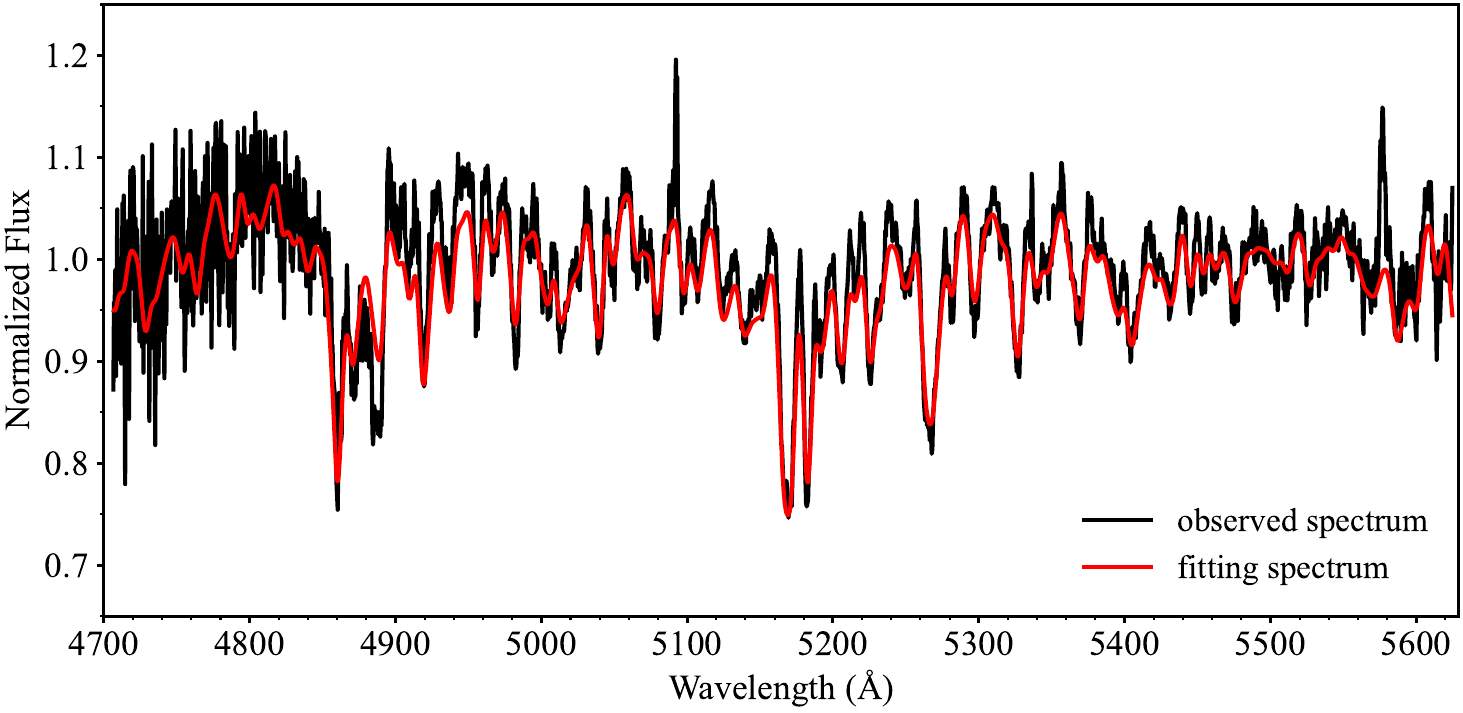}
\caption{Spectra of BFOSC near phase 0. The black solid line is the observed spectra, while the red solid line is the fitting spectra obtained by ULySS.}
\label{f-ULYSS}
\end{figure*}

\begin{figure*}
\centering
\includegraphics[scale=0.45]{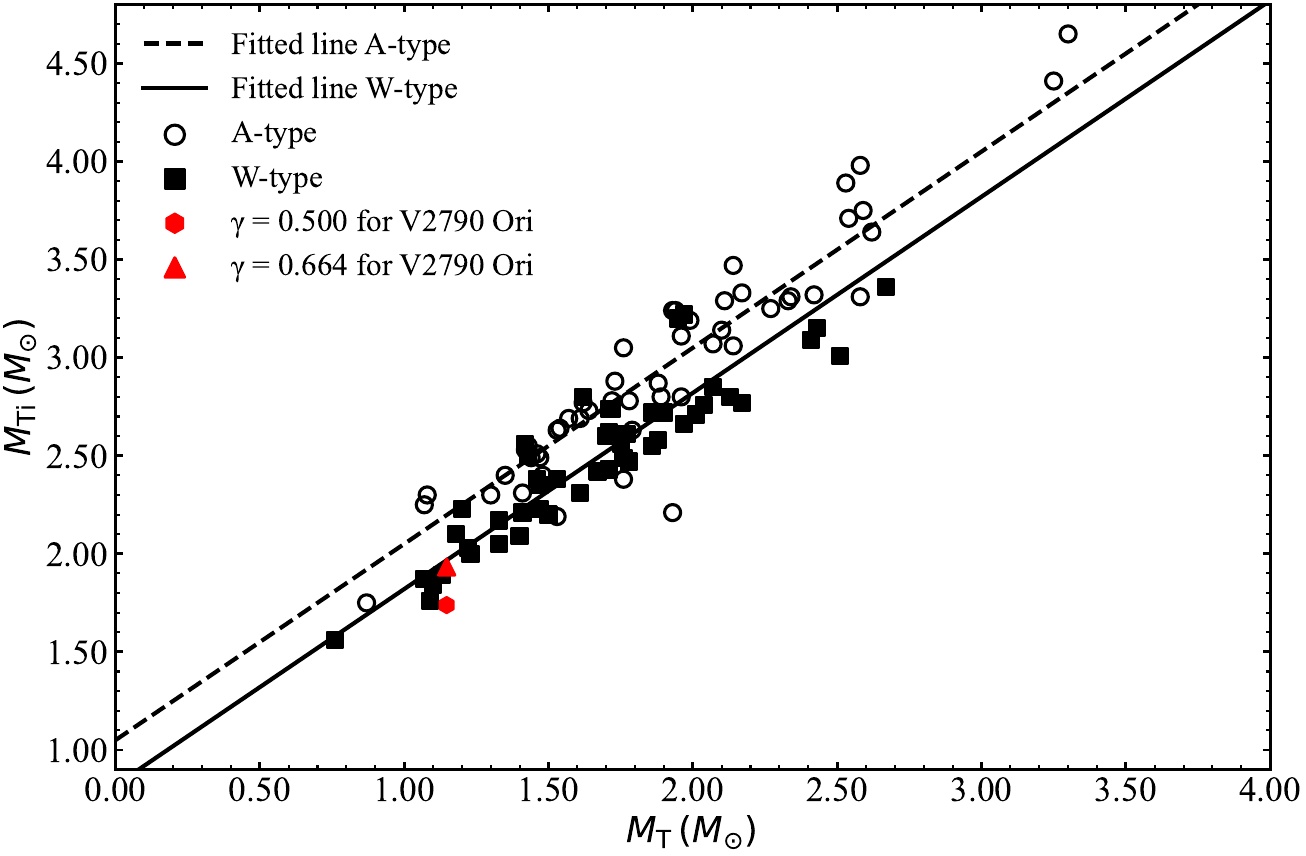}
\caption{The relation between present total mass $M_{\mathrm{T}}$ and initial total mass $M_{\mathrm{Ti}}$ for A-subtype and W-subtype contact binaries from \citet{2014MNRAS.437..185Y} using a sample of 51 A-subtype (open circles) and 49 W-subtype (solid circles) contact systems, respectively. The position of V2790~Ori ($\gamma = 0.597$) is marked with a red solid star.}
\label{f-MM}
\end{figure*}

\citet{2025ApJ...986...25G}This Appendix provides supplementary tables and figures for V2790~Ori. Table~\ref{ta1} presents the photometric data of V2790~Ori. Figure~\ref{f-ULYSS} displays the spectra of BFOSC near phase 0. Table~\ref{ta2} provides the radial velocities of V2790~Ori. Table~\ref{t-spot} lists the spot parameters of V2790~Ori. Figure~\ref{f-MM} shows the relation between present total mass and initial total mass. Table~\ref{t-mini} gives the eclipsing minima of V2790~Ori. All calculated measurements of the O’Connell effect in the light curves of V2790~Ori are summarized in Table~\ref{t-OER}.

\tabletypesize{\scriptsize}
\begin{deluxetable*}{cccccc}
\tablecaption{Photometric Solutions and Fitting Results of $O-C$ for V2790~Ori from Previous and This Studies \label{t-previous}}
\tablewidth{0pt}
\tablehead{
\multicolumn{6}{c}{\text{Photometric Solutions}} \\
\hline
\colhead{Parameters} & \colhead{\citet{2016JAVSO..44...30M}} & 
\colhead{\citet{2019RAA....19..143K}} & \colhead{\citet{2020NewA...8001400S}} & 
\colhead{\citet{2021AcA....71..123A}} & \colhead{This study}\\
& (AAVSO 2015) & (TNO 2015) & (KAO 2017) & (DBO 2020) & 
}                                     
\startdata
$q$                                       & 0.317(1)          & 0.341(0)       & 0.312(0)      & 0.304(1)          & 0.322(3)                          \\
$T_1(K)$                                  & 5471              & 5644           & 5446(12)      & 5590              & 5314                          \\
$T_2(K)$                                  & 5620(4)           & 5856(9)        & 5567          & 5802(2)           & 5489(2)                          \\
$i(^{\circ})$                             & 84.15(20)         & 85.1(2)        & 82.7(2)       & 84.93(19)         & 84.3(2)                          \\
$L_{2B}/(L_{1B}+L_{2B})$                  & ---               & 0.3229(10)     & ---           & 0.3025(2)         & 0.3105(6)                     \\
$L_{2V}/(L_{1V}+L_{2V})$                  & ---               & 0.3106(7)      & 0.277(25)     & 0.2889(1)         & 0.2993(4)                     \\
$L_{2R_{C}}/(L_{1R_{C}}+L_{2R_{C}})$      & ---               & 0.3048(6)      & 0.274(14)     & ---               & 0.2934(4)                     \\
$L_{2I_{C}}/(L_{1I_{C}}+L_{2I_{C}})$      & ---               & ---            & 0.284(30)     & 0.2784(1)         & 0.2892(3)                    \\
$L_{2g'}/(L_{1g'}+L_{2g'})$               & 0.2966(7)         & ---            & ---           & ---               & 0.3054(5)                          \\
$L_{2r'}/(L_{1r'}+L_{2r'})$               & 0.2867(5)         & ---            & ---           & ---               & 0.2943(4)                         \\ 
$L_{2i'}/(L_{1i'}+L_{2i'})$               & 0.2830(5)         & ---            & ---           & ---               & 0.2899(3)                         \\   
$L_{2T}/(L_{1T}+L_{2T})$                  & ---               & ---            & ---           & ---               & 0.2898(3)                          \\
$r_1$                                     & ---               & 0.4865(99)     & 0.4854        & 0.4878(2)         & 0.4861(2)                           \\   
$r_2$                                     & ---               & 0.3011(133)    & 0.2693        & 0.2824(2)         & 0.2928(4)                  \\               
$f(\%)$                                   & 15(2)             & 20.89(1.03)    & 12.8(1.6)     & 5.0(5)            & 14.8(0.6)                      \\
\hline
Spot 1                                    & Cool spot         & ---           & Hot spot      & ---                & ---                         \\                     
\hline
Colatitude ($^{\circ}$)                   & 78(4)             & ---           & 90            & ---                & ---                         \\        
Longitude ($^{\circ}$)                   & 2(1)              & ---           & 80            & ---                & ---                         \\
Angular Radius ($^{\circ}$)               & 12(4)             & ---           & 11            & ---                & ---                         \\
T-factor                                  & 0.90(5)           & ---           & 1.2           & ---                & ---                         \\
\hline                                                                                                                                     
Spot 2                                    & Hot spot          & Cool spot      & ---          & Cool spot          & ---                           \\                     
\hline
Colatitude ($^{\circ}$)                   & 105(5)            & 37(3)          & ---          & 90(4)              & ---                           \\        
Longitude ($^{\circ}$)                   & 10(3)             & 264(4)         & ---          & 179.5(5)           & ---                           \\
Angular Radius ($^{\circ}$)               & 14(4)             & 26(2)          & ---          & 12.0(1)            & ---                            \\
T-factor                                  & 1.16(5)           & 0.90(3)        & ---          & 0.75(1)            & ---                            \\
\hline
\multicolumn{6}{c}{\text{Fitting Results of $O-C$}} \\[2pt]
\hline 
$\Delta T_0$ (days)                       & ---               & $-5(\pm21)\times10^{-5}$         & ---     & ---                                & $1.25(\pm1.80)\times10^{-4}$           \\
$\Delta P_0$                              & ---               & $1.67(\pm2.45)\times10^{-7}$     & ---     & ---                                & $-6.44(\pm2.34)\times10^{-8}$           \\
$\dot P$ (d\,yr$^{-1}$)                   & ---               & $1.03(\pm1.43)\times10^{-7}$     & ---     & $-2.22(\pm0.01)\times10^{-8}$      & $-3.18(\pm0.75)\times10^{-8}$           \\
$A$ (days)                                & ---               &  ---                             & ---     & $1.7(\pm0.6)\times10^{-3}$          & $8.98(\pm2.19)\times10^{-4}$            \\
$\omega$ ($^{\circ}$)                     & ---               &  ---                             & ---     & ---                                & ---                                        \\
$P_3$ (yr)                                & ---               &  ---                             & ---     & $6.32(\pm0.16)$                    & $7.44(\pm0.52)$                          \\
$T_3$                                     & ---               &  ---                             & ---     & ---                                & ---                                      \\
$\phi$ ($^{\circ}$)                       & ---               & ---                              & ---     & ---                                & $256.26(\pm11.01)$                       \\
\enddata            
\tablecomments{The table summarizes the photometric solutions and orbital period variation analyses of V2790~Ori reported in previous studies, together with the results obtained in this work for comparison. 
Note that \citet{2016JAVSO..44...30M} and \citet{2020NewA...8001400S} did not perform orbital period variation studies, so their results are not included in the table.}
\end{deluxetable*}

\begin{deluxetable*}{ccccccccc}  
\tabletypesize{\small}
\tablecaption{Photometric Data of V2790~Ori (Our Observations) \label{ta1}}
\tablewidth{0pt}
\tablehead{
\colhead{LCs} & \colhead{Time (BJD)} & \colhead{$\Delta m_{B}$} & 
\colhead{Time (BJD)} & \colhead{$\Delta m_{V}$} &
\colhead{Time (BJD)} & \colhead{$\Delta m_{R_{C}}$} &
\colhead{Time (BJD)} & \colhead{$\Delta m_{I_{C}}$}
}
\startdata
2015 Feb 05 \& Mar 06   &   2457058.93893   & 0.326 & 2457058.93932    & 0.490  & 2457058.93971    & 0.569  & 2457058.93997    & 0.604  \\
                         &   2457058.94038   & 0.316 & 2457058.94077    & 0.479  & 2457058.94116    & 0.552  & 2457058.94142    & 0.598  \\
                         &   2457058.94183   & 0.309 & 2457058.94221    & 0.466  & 2457058.94261    & 0.547  & 2457058.94287    & 0.587  \\
                         &   2457058.94328   & 0.297 & 2457058.94367    & 0.461  & 2457058.94405    & 0.538  & 2457058.94431    & 0.579  \\
                         &   2457058.94472   & 0.293 & 2457058.94511    & 0.453  & 2457058.94550    & 0.528  & 2457058.94576    & 0.571  \\
                         &   2457058.94617   & 0.278 & 2457058.94656    & 0.448  & 2457058.94695    & 0.524  & 2457058.94721    & 0.563  \\
                         &   2457058.94762   & 0.277 & 2457058.94801    & 0.438  & 2457058.94898    & 0.509  & 2457058.94980    & 0.551  \\
                         &   2457058.95021   & 0.261 & 2457058.95060    & 0.426  & 2457058.95132    & 0.500  & 2457058.95158    & 0.545  \\
                         &   2457058.95199   & 0.257 & 2457058.95238    & 0.419  & 2457058.95276    & 0.496  & 2457058.95302    & 0.539  \\
                         &   2457058.95343   & 0.248 & 2457058.95382    & 0.415  & 2457058.95420    & 0.490  & 2457058.95446    & 0.532  \\
\ldots & \ldots & \ldots & \ldots & \ldots & \ldots & \ldots & \ldots & \ldots \\
\enddata
\tablecomments{This table is available in its entirety in machine-readable form in the online journal. A portion is shown here for guidance regarding its form and content.}
\end{deluxetable*}

\begin{deluxetable*}{cccccc}  
\tablecaption{Radial Velocities of V2790~Ori 
\label{ta2}}
\tablewidth{0pt}
\tablehead{
\colhead{Time (BJD)} & \colhead{Phase} & \colhead{$\mathrm{RV}_{\mathrm{L}}$} & 
\colhead{Errors} & 
\colhead{$\mathrm{RV}_{\mathrm{M}}$} &
\colhead{Errors} 
}
\startdata
2458848.17407 & 0.18026  & ---        & ---    & 0.7     & 1.7   \\
2458848.18456 & 0.21669  & -256.4     & 16.0   & 32.8    & 11.5  \\
2458848.19505 & 0.25312  & -326.2     & 5.6    & -1.0    & 5.8   \\
2458848.26667 & 0.50194  & -50.6      & 3.1    & ---     & ---   \\
2458848.30469 & 0.63403  & 141.9      & 8.0    & -105.9  & 6.5   \\
2458848.31517 & 0.67046  & 177.6      & 20.5   & -133.3  & 13.1  \\
2458848.32566 & 0.70689  & 181.5      & 11.1   & -94.0   & 9.3   \\
2458848.33614 & 0.74332  & 204.9      & 14.4   & -120.2  & 9.0   \\
2459536.13018 & 0.23115  & -313.6     & 2.3    & 9.8     & 1.6   \\
2459536.14069 & 0.26937  & -305.0     & 2.9    & 10.8    & 2.0   \\
2459536.15120 & 0.30411  & -292.4     & 3.2    & 15.6    & 2.1   \\
2459536.16170 & 0.34233  & -266.8     & 4.8    & 9.6     & 3.5   \\
2459536.17221 & 0.37707  & -224.8     & 8.2    & 20.2    & 6.0   \\
2459536.18270 & 0.41528  & -202.6     & 10.6   & 5.6     & 7.8   \\
2459536.19321 & 0.45002  & ---        & ---    & -27.5   & 1.2   \\
2459536.29315 & 0.79744  & 199.2      & 4.2    & -105.3  & 2.7   \\
2459536.30366 & 0.83565  & 187.4      & 4.9    & -103.4  & 3.1   \\
2459536.31422 & 0.87039  & 144.1      & 6.7    & -109.1  & 10.7  \\
2459536.32473 & 0.90861  & 132.5      & 5.9    & -84.8   & 4.4   \\
2459536.33524 & 0.94335  & ---        & ---    & -54.4   & 1.2   \\
\enddata
\tablecomments{$\mathrm{RV}_{\mathrm{L}}$ represents the radial velocities of the less massive components, while $\mathrm{RV}_{\mathrm{M}}$ represents the radial velocities of the more massive components.}
\end{deluxetable*}

\begin{deluxetable*}{cccccccccc}
\tablecaption{Spot Parameters of V2790~Ori \label{t-spot}}
\tablewidth{0pt}
\tablehead{
\colhead{LCs} &
\multicolumn{3}{c}{Spot 1} & \multicolumn{3}{c}{Spot 2} \\
\cline{2-7}
\colhead{} & 
\colhead{Longitude ($^{\circ}$)} & 
\colhead{Angular Radius ($^{\circ}$)} & 
\colhead{T-factor} &
\colhead{Longitude ($^{\circ}$)} & 
\colhead{Angular Radius ($^{\circ}$)} & 
\colhead{T-factor}
}
\startdata
2015 Feb 05 \& Mar 06              & 183.4      & 18.1     & 0.726        & 158.8    & 8.2    & 0.718  \\
2015 Dec 28                        & 354.1      & 11.8     & 0.726        & 165.0    & 9.1    & 0.718  \\
2016 Dec 30                        & 130.3      & 24.3     & 0.728        & 186.9    & 10.0   & 0.704  \\
2017 Dec 20                        & 178.8      & 22.0     & 0.717        & 167.2    & 10.0   & 0.658  \\ 
2018 Dec 27                        & 320.4      & 5.2      & 0.728        & 263.7    & 5.6    & 0.728  \\ 
2019 Dec 29 \& 2020 Jan 03         & 328.5      & 18.6     & 0.726        & 258.0    & 13.0   & 0.726  \\          
2020 Dec 20                        & 123.2      & 9.6      & 0.898        & 263.4    & 11.7   & 0.580  \\ 
2021 Jan 04 \& Jan 05              & 332.7      & 5.7      & 0.726        & 98.7     & 6.0    & 0.726  \\ 
2022 Jan 02                        & 283.9      & 11.1     & 0.726        & 281.5    & 10.4   & 0.726  \\   
2022 Dec 31                        & 330.0      & 9.9      & 0.690        & 354.4    & 7.5    & 0.631  \\ 
2024 Feb 27 \& Mar 08              & 173.8      & 9.0      & 0.724        & 278.7    & 8.7    & 0.724  \\ 
2025 Jan 08                        & 146.5      & 12.0     & 0.695        & 170.8    & 7.2    & 0.684  \\                       
AAVSO -- 2015 Jan 08 -- Jan 29     & 173.1      & 18.1     & 0.768        & 164.9    & 8.1    & 0.768  \\ 
AAVSO -- 2015 Nov 19 \& Nov 20     & 177.1      & 20.0     & 0.707        & 167.5    & 8.7    & 0.635  \\ 
AAVSO -- 2020 Jan 29 -- Feb 05     & 198.9      & 11.0     & 0.645        & 180.7    & 13.3   & 0.686  \\   
TESS -- S43                        & 30.0       & 14.0     & 0.726        & 334.0    & 14.1   & 0.726  \\
TESS -- S44                        & 32.4       & 8.4      & 0.704        & 314.9    & 13.8   & 0.740  \\
TESS -- S45                        & 31.0       & 11.7     & 0.728        & 327.7    & 12.8   & 0.728  \\
TESS -- S71                        & 340.8      & 18.0     & 0.726        & 282.8    & 13.4   & 0.726  \\
TESS -- S72                        & 346.4      & 19.3     & 0.738        & 277.6    & 14.3   & 0.732  \\
TNO -- 2015 Jan 21 -- Jan 23       & 190.9      & 19.6     & 0.756        & 176.5    & 12.0   & 0.742  \\  
KAO -- 2017 Jan 21 \& Jan 22       & 119.2      & 20.5     & 0.742        & 339.7    & 9.0    & 0.695  \\
\enddata
\tablecomments{Spot 1 is on the primary component, and Spot 2 is on the secondary component.}
\end{deluxetable*}

\begin{deluxetable*}{ccccccc}
\tablecaption{Eclipsing Minima of V2790~Ori \label{t-mini}}
\tablewidth{0pt}
\tablehead{
\colhead{BJD} &
\colhead{Error} &
\colhead{Epoch} &
\colhead{$(O-C)_{1}$} &
\colhead{$(O-C)_{2}$} &
\colhead{Res.} &
\colhead{Ref.}
}
\startdata
2451521.6950 & 0.0005 & -24175   & -0.0044 & 0.0012 & 0.0005 & 1 \\ 
2453327.7560 & 0.0004 & -17900.5 & -0.0034 & -0.0007 & 0.0002 & 2 \\ 
2455520.8205 & 0.0002 & -10281.5 & -0.0015 & -0.0009 & -0.0003 & 3 \\ 
2455532.9103 & 0.0004 & -10239.5 & -0.0010 & -0.0005 & 0.0002 & 2 \\ 
2455604.2950 & 0.0004 & -9991.5  & -0.0009 & -0.0004 & 0.0003 & 3 \\ 
2455632.3597 & 0.0004 & -9894    & -0.0007 & -0.0003 & 0.0005 & 3 \\ 
2455644.3050 & 0.0004 & -9852.5  & -0.0009 & -0.0004 & 0.0004 & 3 \\ 
2455896.8827 & 0.0003 & -8975    & -0.0039 & -0.0036 & -0.0027 & 4 \\ 
2455902.7850 & 0.0001 & -8954.5  & -0.0023 & -0.0020 & -0.0011 & 5 \\ 
2455959.3466 & 0.0002 & -8758    & -0.0015 & -0.0013 & -0.0004 & 6 \\ 
2456288.0610 & 0.0004 & -7616    & -0.0018 & -0.0017 & -0.0012 & 7 \\ 
2456288.2045 & 0.0004 & -7615.5  & -0.0023 & -0.0022 & -0.0017 & 7 \\ 
2456623.1092 & 0.0004 & -6452    & -0.0009 & -0.0009 & -0.0011 & 8 \\ 
2456623.2547 & 0.0004 & -6451.5  & 0.0007  & 0.0007  & 0.0005  & 8 \\ 
2457030.6939 & 0.0004 & -5036    & 0.0007  & 0.0005  & -0.0003 & 9 \\ 
\ldots & \ldots & \ldots & \ldots & \ldots & \ldots & \ldots \\
\enddata
\tablecomments{Ref. 1. \citet{2004IBVS.5570....1O}; 2. \citet{2011IBVS.5960....1D}; 3. \citet{Nagai2012VSOLJ}; 4. \citet{Diethelm2012IBVS6011}; 5. \citet{2012IBVS.6018....1N}; 6. \citet{2013IBVS.6084....1H}; 7. \citet{Nagai2013}; 8. \citet{Hirosawa2013}; 9. AAVSO; 10. \citet{2019RAA....19..143K}; 11. Our data; 12. \citet{2017OEJV..179....1J}; 13. \citet{Nelson2016}; 14. ASAS-SN; 15. \citet{Nagai2016}; 16. \citet{2020NewA...8001400S}; 17. BRNO; 18. TESS; 19. \citet{Nagai2023};
This table is available in its entirety in machine-readable form in the online journal. A portion is shown here for guidance regarding its form and content.}
\end{deluxetable*}


\begin{deluxetable*}{lccccccc}
\tablecaption{Measurements of the O'Connell Effect in the Light Curves of V2790~Ori \label{t-OER}}
\tablewidth{0pt}
\tablehead{
\colhead{LCs} & \colhead{Mean Epoch} & \colhead{Band} & \colhead{$\Delta m$ (mag)} & \colhead{OER} & \colhead{Min.I - Min.II} & \colhead{Max.I - Min.II} & \colhead{Max.II - Min.II}
}
\startdata
2015 Feb 05 \& Mar 06     & -4837     & $B$       & -0.0051  & 0.9796   & 0.0132	 & -0.6074 	 & -0.6125 \\                                                                                                         
                          &           & $V$       & -0.0054  & 0.9801   & 0.0053	 & -0.5651	 & -0.5706 \\                                                                                                        
                          &           & $R_{C}$   & -0.0055  & 0.9812   & 0.0035	 & -0.5399	 & -0.5454 \\                                                                                                         
                          &           & $I_{C}$   & -0.0054  & 0.9826   & 0.0058	 & -0.5169	 & -0.5222 \\     
2015 Dec 28               & -3804     & $B$       & -0.0062	 & 0.9806	  & 0.0474	 & -0.5704	 & -0.5765 \\   
                          &           & $V$       & -0.0066	 & 0.9805	  & 0.0391	 & -0.5342	 & -0.5407 \\
                          &           & $R_{C}$   & -0.0067	 & 0.9812	  & 0.0348	 & -0.5136	 & -0.5203 \\
                          &           & $I_{C}$   & -0.0066	 & 0.9823	  & 0.0340	 & -0.4949	 & -0.5016 \\   
2016 Dec 30               & -2626     & $B$       & 0.0460	 & 1.1035	  & -0.0020	 & -0.6073	 & -0.5613 \\   
                          &           & $V$       & 0.0421	 & 1.0986	  & -0.0028	 & -0.5687	 & -0.5266 \\
                          &           & $R_{C}$   & 0.0386	 & 1.0924	  & -0.0024	 & -0.5453	 & -0.5067 \\
                          &           & $I_{C}$   & 0.0350	 & 1.0852	  & 0.0009	 & -0.5236	 & -0.4886 \\  
2017 Dec 20               & -1293     & $B$       & -0.0018	 & 0.9868	  & -0.0003	 & -0.5902	 & -0.5971 \\ 
                          &           & $V$       & -0.0022	 & 0.9867	  & -0.0044	 & -0.5521	 & -0.5558 \\
                          &           & $R_{C}$   & -0.0024	 & 0.9871	  & -0.0045	 & -0.5299	 & -0.5308 \\
                          &           & $I_{C}$   & -0.0026	 & 0.9878	  & -0.0011	 & -0.5099	 & -0.5077 \\
2018 Dec 27               & 0         & $B$       & 0.0063	 & 1.0087	  & 0.0776	 & -0.5692	 & -0.5630  \\
                          &           & $V$       & 0.0058	 & 1.0093	  & 0.0552	 & -0.5369	 & -0.5311  \\
                          &           & $R_{C}$   & 0.0052	 & 1.0092	  & 0.0456	 & -0.5184	 & -0.5132  \\
                          &           & $I_{C}$   & 0.0046	 & 1.0089	  & 0.0420	 & -0.5011	 & -0.4965  \\   
 \ldots & \ldots & \ldots & \ldots & \ldots & \ldots & \ldots & \ldots \\
\enddata                     
\tablecomments{This table is available in its entirety in machine-readable form in the online journal. A portion is shown here for guidance regarding its form and content.}
\end{deluxetable*}

\bibliography{sample701}{}
\bibliographystyle{aasjournalv7}



\end{document}